\documentclass[printer]{aa}

\usepackage[utf8]{inputenc}
\usepackage{graphicx}
\usepackage{xcolor}
\usepackage[colorlinks=true,breaklinks=true,citecolor=blue,linkcolor=blue,urlcolor=blue]{hyperref}

\begin{document}

\title{Dust entrainment in photoevaporative winds:\\Densities and imaging}

\author{R.~Franz \inst{1}\thanks{\href{mailto:rfranz@usm.lmu.de}{rfranz@usm.lmu.de}} \and B.~Ercolano \inst{1,2} \and S.~Casassus \inst{3} \and G.~Picogna \inst{1} \and T.~Birnstiel \inst{1,2} \and S.~P{\'e}rez \inst{4,5} \and Ch.~Rab \inst{1,6} \and A.~Sharma \inst{1}}

\institute{University Observatory, Faculty of Physics, Ludwig-Maximilians-Universit{\"a}t M{\"u}nchen, Scheinerstr.~1, 81679 Munich, Germany
\and
Excellence Cluster Origin and Structure of the Universe, Boltzmannstr.~2, 85748 Garching, Germany
\and
Departamento de Astronom{\'i}a, Universidad de Chile, Casilla 36-D, Santiago, Chile
\and
Departamento de F{\'i}sica, Universidad de Santiago de Chile, Av.~Ecuador 3493, Estaci{\'o}n Central, Santiago, Chile
\and
Center for Interdisciplinary Research in Astrophysics and Space Exploration (CIRAS), Universidad de Santiago de Chile, Chile
\and
Max-Planck-Institut f{\"u}r extraterrestrische Physik, Giessenbachstr.~1, 85748 Garching, Germany
}

\date{Received 15 Mar 2021 / Accepted 18 Oct 2021}

\abstract
{X-ray- and extreme-ultraviolet- (together: XEUV-) driven photoevaporative winds acting on protoplanetary disks around young T-Tauri stars may crucially impact disk evolution, affecting both gas and dust distributions.}
{We constrain the dust densities in a typical XEUV-driven outflow, and determine whether these winds can be observed at $\mu$m-wavelengths.}
{We used dust trajectories modelled atop a 2D hydrodynamical gas model of a protoplanetary disk irradiated by a central T-Tauri star.
With these and two different prescriptions for the dust distribution in the underlying disk, we constructed wind density maps for individual grain sizes.
We used the dust density distributions obtained to synthesise observations in scattered and polarised light.}
{For an XEUV-driven outflow around a $M_* = 0.7\,\mathrm{M}_\odot$ T-Tauri star with $L_X = 2 \cdot 10^{30}\,\mathrm{erg/s}$, we find a dust mass-loss rate $\dot{M}_\mathrm{dust} \lesssim 4.1 \cdot 10^{-11}\,\mathrm{M_\odot / yr}$ for an optimistic estimate of dust densities in the wind (compared to $\dot{M}_\mathrm{gas} \approx 3.7 \cdot 10^{-8}\,\mathrm{M_\odot / yr}$).
The synthesised scattered-light images suggest a distinct chimney structure emerging at intensities $I/I_{\max} < 10^{-4.5}$ ($10^{-3.5}$) at $\lambda_\mathrm{obs} = 1.6$ (0.4) $\mu$m, while the features in the polarised-light images are even fainter.
Observations synthesised from our model do not exhibit clear features for SPHERE IRDIS, but show a faint wind signature for JWST NIRCam under optimal conditions.}
{Unambiguous detections of photoevaporative XEUV winds launched from primordial disks are at least challenging with current instrumentation; this provides a possible explanation as to why disk winds are not routinely detected in scattered or polarised light.
Our calculations show that disk scale heights retrieved from scattered-light observations should be only marginally affected by the presence of an XEUV wind.}


\maketitle


\section{Introduction}
\label{sec:Intro}

Planet formation needs large amounts of material and is thus expected to happen in and close to the midplane of protoplanetary disks around young stars \citep[see e.g.][]{Armitage-2010}.
Its outcome may be affected by disk winds, which remove material from the disk, and may thus change the final masses and locations of planets by altering the dust-to-gas ratio and halting inward migration \citep[see e.g.][]{Alexander-2012a, Ercolano-2015a, Ercolano-2017c, Carrera-2017, Jennings-2018, Monsch-2019, Monsch-2021b}.

While current literature agrees on two major scenarios that can drive disk winds -- that is magnetics- and photoevaporation-driven outflows \citep[see e.g.][]{Owen-2010, Gressel-2015, Bai-2016b, Wang-2019, Picogna-2019, Woelfer-2019, Rodenkirch-2020} -- it remains unclear which mechanism dominates during which evolutionary stages \citep[see e.g.][]{Ercolano-2017b, Coutens-2019, Gressel-2020}.
Consequently, many efforts have been focussed on determining which varieties of winds can be encountered in protoplanetary disks.
One approach is to look for and model wind signatures in gas tracers \citep[e.g.][for some recent works]{Pascucci-2020, Gangi-2020, Weber-2020}.
However, gas tracers identified so far are difficult to interpret and cannot be unambigously inverted to obtain mass-loss rates from the winds \citep[e.g.][]{Ercolano-2016}; hence, forays to quantify dust entrainment in winds have recently gained traction \citep{Hutchison-2016b, Jaros-2020, Vinkovic-2021} despite their high computational cost \citep[see e.g.][]{Tamfal-2018}.

\citet{Miyake-2016} have shown that magneto-rotational turbulences can support an outflow of dust grains of several tens of $\mu$m from the inner disk region.
Observationally, dust grains above the midplane of protoplanetary disks have been suggested mainly for regions close to the star, where high densities may lead to a significant shadowing of the stellar luminosity \citep{Dodin-2019, Petrov-2019}.\footnote{\citet{Garate-2019} have presented an additional theoretical model to explain these observations.}

Dust entrainment in photoevaporative extreme-ultraviolet- (EUV-) driven winds has first been investigated by \citet{Owen-2011a}, later by \citet{Hutchison-2016c, Hutchison-2016b}, and very recently by \citet{Hutchison-2021} and \citet{Booth-2021}, building on the semi-analytical modelling of \citet{Clarke-2016} \citep[which has recently been generalised by][]{Sellek-2021}.
All studies found an entrainment of dust grains of several $\mu$m in size.
In an earlier work \citep[][henceforth Paper I]{Franz-2020}, we studied dust motion in X-ray-driven winds based on a new generation of X-ray and EUV (together: XEUV) gas models developed by \citet{Picogna-2019}, and found dust particles up to about 10\,$\mu$m to be entrained, moving on a much faster timescale than in the disk midplane \citep[see][]{Misener-2019}.

Despite the evident interest in the subject, the question of whether dust outflows driven by photoevaporation are currently observable has remained unanswered to date.
At the time of writing, no firm detection of such a dusty wind component exists in the literature, but this might be due to the sample selection.
Thus, in this work, we aim to provide a theoretical prediction of the observability of this wind component, which may guide future observational campaigns and help interpret the datasets that are currently available.

This paper is organised as follows: The calculations to obtain the dust densities and synthetic observations are outlined in Sect.~\ref{sec:Methods}.
In Sect.~\ref{sec:Results}, the resulting density maps and observational concurrences are shown.
We discuss our findings in Sect.~\ref{sec:Discussion} and summarise them in Sect.~\ref{sec:Summary}.


\section{Methods}
\label{sec:Methods}

\citet{Picogna-2019} have provided models of protoplanetary gas disks with a photoevaporative XEUV wind, computed using \texttt{Mocassin} \citep{Ercolano-2003, Ercolano-2005, Ercolano-2008a} for the radiative transfer computations,\footnote{\texttt{Mocassin}: \href{https://mocassin.nebulousresearch.org/}{[link]}} and \texttt{Pluto} \citep{Mignone-2007} for the successive hydrodynamical (HD) evolution.\footnote{\texttt{Pluto}: \href{http://plutocode.ph.unito.it/}{[link]}. Version 4.2 was used for this work.}
In Paper I, we have used such a gas disk to simulate a collection of dust grain trajectories in its wind region, and we refer to that work for details about the modelling set-up. We now proceed to use these individual trajectories for the creation of dust density maps and successively, synthetic observations.

\subsection{Dust densities in the XEUV wind}
\label{sec:Methods:rho}

The gas disk employed in Paper I models a $M_\mathrm{disk} \simeq 0.01\,M_*$ primordial disk around a young $M_*=0.7\,\mathrm{M}_\odot$ T-Tauri star with an X-ray luminosity of $L_X = 2 \cdot 10^{30}\,\mathrm{erg/s}$ \citep{Preibisch-2005},\footnote{When using the term primordial disk in this work, we mean a primordial Class-II disk (i.e. without a cavity).} driving a gas mass-loss rate of $\dot{M}_\mathrm{gas} \simeq 3.7 \cdot 10^{-8}\,\mathrm{M}_\odot$/yr \citep{Picogna-2019}.
The $\dot{M}_\mathrm{gas}$ encountered in this one gas disk snapshot is somewhat higher than the $2.6 \cdot 10^{-8}\,\mathrm{M}_\odot$/yr given in \citet{Picogna-2019} because in the latter work, a time-averaged value was used.

The dust grain motion was modelled on top of this gas snapshot using the Lagrangian-particle approach of \citet{Picogna-2018}; this yielded a collection of individual grain trajectories for a variety of grain sizes (i.e. radii) $a_0$ in the XEUV-driven wind region of the disk.

\subsubsection{Dust grains: From trajectories to distributions}
\label{sec:Methods:rho:traj-dist}

As detailed in Paper I, these dust grains were launched from the disk surface into a 2D gas disk with 3D velocity information.
They are accelerated by the gravity of the central star and the gas drag, and may experience turbulent kicks while within the disk; disk gravity was shown to be negligible in Paper I.

As expected, small dust grains were found to be mostly entrained in the gas flow, whereas larger ones decoupled from it; the largest grains entrained by the wind (for an internal grain density of $\varrho_\mathrm{grain} = 1\,\mathrm{g/cm^3}$) had a size $a_0 = 11\,\mu\mathrm{m}$ and a Stokes number $St < 0.4$ when initially lifted up.
So in order to sensibly model dust densities, we chose to create density maps for a discrete sample of $a_0$, that is $a_0 \in \lbrace 0.01, 0.1, 0.5, 1, 2, 4, 8, 10 \rbrace \,\mu$m; this limitation to eight discrete values is due to the high computational cost for simulating the trajectories.
From a physical point of view, the listed sizes cover all major cases between full entrainment, slow decoupling and fast decoupling (for which $a_0 \in \lbrace 0.1, 4, 10 \rbrace \,\mu$m were showcased in Paper I).

As listed in Table~\ref{tab:particle-counts}, we have modelled at least 200\,000 dust grain trajectories for each of the $a_0$ investigated in order to provide reasonably high spatial resolution.
The grains were initially positioned along the disk surface, using a uniform distribution within $0.33 < r\,[\mathrm{AU}] < 200$, with $r=\sqrt{x^2+y^2+z^2}$, yielding more than 1000 grains per radial AU of disk surface for each $a_0$.\footnote{For a more comprehensive take on the disk surface, see Paper I; in short, it is defined as the location of the largest drop in gas temperature.}
The varying amounts of particles modelled are due to the fixed simulation time of $\Delta t \simeq 3.8\,$kyr (corresponding to one orbit at 200\,AU).\footnote{Doubling $\Delta t$ results in very sparse, relative local errors $<25$\% (mostly $\ll 10$\%) for the dust densities. This error is negligible compared to the uncertainties in the base density estimation of Sect.~\ref{sec:Methods:rho:base}.}
For all $a_0$, the concurrent amount of particles in the simulation is 200\,000; when a particle exits the computational domain ($0.33 \leq r\,[\mathrm{AU}] \leq 300$ and $0.001 \leq \vartheta\,[\mathrm{rad}] \leq \pi/2$), it is re-inserted at a random position along the disk surface, starting an additional trajectory.
Thus, we obtained more trajectories for smaller grains which move faster.

\begin{table}
    \centering
    \caption{Statistics for the dust trajectories used to create the density maps: grain size, number of all trajectories modelled, number of fully entrained trajectories thereof.}
    \begin{tabular}{crr}
        \hline\hline
        $a_0$ [$\mu$m] & \multicolumn{1}{l}{$N_\mathrm{all}$} & \multicolumn{1}{l}{$N_\mathrm{entrained}$} \\ \hline
        0.01 & 5\,495\,998 & 4\,497\,540 \\
        0.1  & 5\,492\,527 & 5\,012\,884 \\
        0.5  & 4\,555\,823 & 4\,329\,718 \\
        1    & 3\,968\,894 & 3\,730\,741 \\
        2    & 3\,138\,379 & 2\,834\,014 \\
        4    & 1\,537\,412 & 1\,274\,661 \\
        8    &    508\,545 &    217\,772 \\
        10   &    337\,515 &     33\,294 \\ \hline
    \end{tabular}
    \tablefoot{The differing numbers stem from a constant sample size of 200\,000 grains processed simultaneously over similar simulation time spans, with grains being reinserted at a random position along the disk surface once they exit the computational domain.
    ($N_\mathrm{entrained}$ is smaller for 0.01\,$\mu$m than for 0.1\,$\mu$m because the former grains are even more coupled to the gas stream which points slightly back towards the disk surface at larger $R$, see Paper I.)}
    \label{tab:particle-counts}
\end{table}

The high entrainment fractions of Table~\ref{tab:particle-counts} match the results of \citet{Hutchison-2021}, who have performed a semi-analytical modelling of dust trajectories below and in an EUV-driven wind, and find entrainment for almost all the grains that could be delivered to the ionisation front.
The fractions for the largest entrainable grains seem to drop, yet these do not necessarily reach the surface area in the first place \citep[see][Paper I, and below]{Hutchison-2016c, Booth-2021}.

The dust density maps were obtained by mapping the dust trajectories to a grid via the particle-in-cell method, applied independently for each $a_0$.
We employed an underlying $(r, \vartheta)$-grid with cell sizes $\Delta r = 1\,$AU ($\Delta r = 0.1\,$AU for $r \leq 2\,$AU) and $\Delta \vartheta = 0.5^{\circ}$.\footnote{As $a_0 \lll \Delta r$ and $a_0 \lll r \, \Delta \vartheta$, no smoothing kernel was applied when mapping the particles onto this grid.}
Dust trajectories were grouped together according to their initial coordinates, and then used to retrieve one dust map per initial $(r, \vartheta)$-bin, yielding a grid with particle counts per cell.
These counts were obtained from the recorded grain positions and velocities, based on a simplified equation of motion:
\begin{equation}
    \vec{r}(t_2) = \vec{r}(t_1) + \frac{1}{2} \, \left( \vec{v}(t_2)+\vec{v}(t_1) \right) (t_2-t_1) \; .
\label{eq:motion}
\end{equation}
The overall contribution of one particle per each (constant) simulation time step to the entire map was normalised to one.
The resulting particle counts (i.e. dust masses) were successively converted to densities by dividing by the 3D cell volumes of the bins, assuming full azimuthal symmetry for the 2D grid.
Subsequently, the base cells were used to normalise the dust flow originating from them.
In a last step, they were weighted by the actual fraction of the disk surface within them.

\subsubsection{Base densities}
\label{sec:Methods:rho:base}

While it is widely assumed that grains of mm-size quickly settle towards the midplane, the vertical mixing of $\mu$m-sized particles is less well-constrained.
For instance, \citet{Pinte-2008} have found a non-negligible signal from $1.6\,\mu$m grains from the surface of the inner disk of IM Lup, yet it is important to note that the signal is significantly weaker already for $\gtrsim 3\,\mu$m.
More recently, the general presence of vertical settling has been verified by a series of ALMA observations \citep[][and many others]{ALMA-2015, Pinte-2016, Villenave-2020}.
Additionally, as pointed out by \citet{Avenhaus-2018} based on a sample of T-Tauri disks observed with SPHERE, protoplanetary disks are in all likelihood rather diverse; vertical mixing may hence vary between individual objects.
This has been corroborated by \citet{Villenave-2019}.
Thus, theoretical estimates range from a globally constant dust-to-gas ratio for $\mu$m-sized grains \citep[e.g.][and references therein]{Takeuchi-2005b} to a strong dependence of the vertical dust scale height on $a_0$ even for small grains \citep[e.g.][]{Dullemond-2004, Fromang-2009, Birnstiel-2010a, Birnstiel-2016, Hutchison-2016b, Hutchison-2018}.

Because the base densities have a considerable impact on the resulting dust density maps in our model, we decided to investigate two different setups.
Firstly, we assumed a fixed dust-to-gas ratio throughout the disk; we refer to this model as `fixed' below.

For the second case, hereafter denoted as `variable', we used a dust scale height prescription providing a rather strong fall-off of the dust densities with $a_0$ over $z$; this was done via the \texttt{disklab} scripts collection (Dullemond \& Birnstiel, in prep.).
In this approach, we used the gas disk data and the same dust-to-gas ratio.
The gas disk was then rendered into hydrostatic equilibrium \citep[see e.g.][]{Armitage-2010} such that
\begin{equation}
    \frac{\partial}{\partial z} \left( c_s^2 \cdot \varrho_\mathrm{gas} \right) = - \varrho_\mathrm{gas} \cdot \frac{G \, M_* \, z}{r^3} \; ,
\label{eq:gas-hs-eq}
\end{equation}
with $\sqrt{P_\mathrm{gas}/\varrho_\mathrm{gas}} = c_s = \sqrt{(k_B \, T) / (\mu \, m_H)}$ the speed of sound and $\mu = 1.37125$ the mean atomic mass in proton masses $m_H$ \citep[same value as in][and Paper I]{Picogna-2019}.
The dust densities were then computed for a vertical settling-mixing equilibrium \citep{Fromang-2009} such that
\begin{equation}
    \frac{\partial}{\partial z} \left( \frac{\varrho_\mathrm{dust}}{\varrho_\mathrm{gas}} \right) = - \frac{\Omega_K^2 \, t_\mathrm{stop}}{D} \cdot z \; ,
\label{eq:dust-vsm-eq}
\end{equation}
with $\Omega_K = \sqrt{G\,M_* / r^3}$, $t_\mathrm{stop}$ the dust grain stopping time, and $D$ the diffusion coefficient \citep[see][and sources within]{Fromang-2009}.

For both the `fixed' and `variable' cases, we employed a MRN distribution \citep{Mathis-1977} with $n(a) \propto a^{-3.5} \, \mathrm{d}a$, using a logarithmically-spaced grid with 400 bins for $1\,\mathrm{nm} = a_{\min} \leq a \leq a_{\max} = 1\,$mm to quantify the relative abundances of dust grain sizes.
The underlying total dust-to-gas (mass) ratio was set to the usually assumed value of 0.01;\footnote{Larger values of the dust-to-gas ratio are possible \citep[see e.g.][]{Miotello-2017, Soon-2019}, and would lead to a higher dust content in the wind.} this yields a dust mass fraction of $\approx 10^{-3}$ for grains $\lesssim 10\,\mu$m in relation to the gas.
A maximum grain size of 1\,mm for the total dust content of the disk is a lower limit \citep[see e.g.][]{Hutchison-2021}, but corresponds to the largest grains proven to exist by ALMA \citep[e.g.][]{ALMA-2015}; hence it serves our intention of providing a maximum estimate, as higher $a_{\max}$ would entail a lower dust content for the wind.

These 400 bins of the MRN distribution were combined into eight bins for the eight grain sizes included in our model.
This was done by fully attributing the contribution of a bin to the (linearly) closest grain size. Contributions from grains $> 11.5\,\mu$m were discarded since these cannot be entrained (see Paper I) and hence cannot populate the wind region.\footnote{The large grains cannot be in the wind region; they could be in the disk region, where they might marginally increase brightness. So disregarding them will either not affect or slightly enhance wind visibility.}

\begin{figure*}
    \centering
    \includegraphics[width=0.495\textwidth]{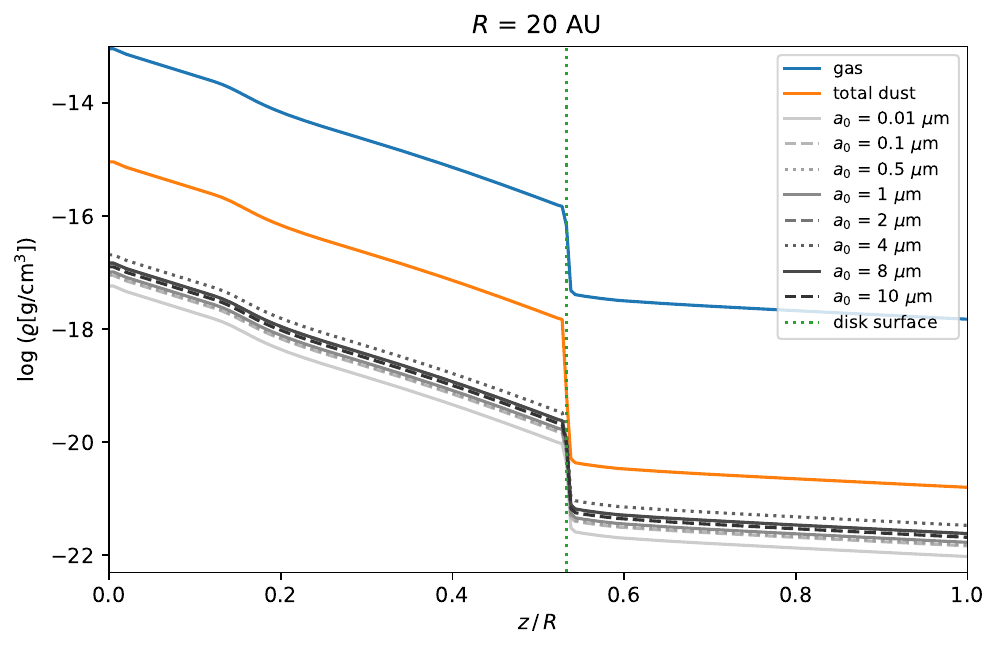}
    \includegraphics[width=0.495\textwidth]{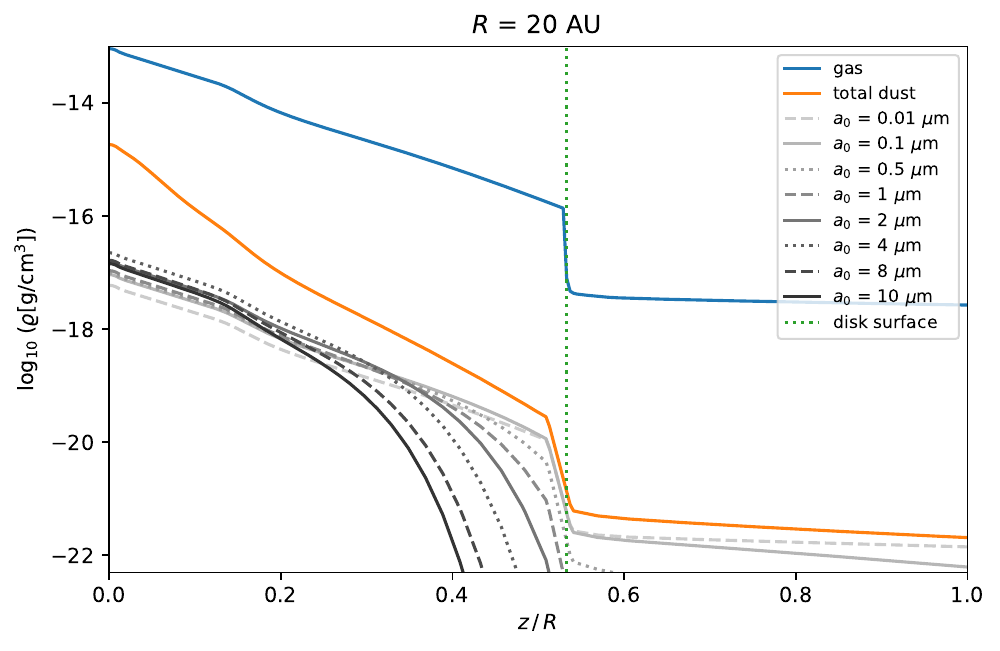}
    \caption{Base density profiles for a vertical slice at $R \approx 20\,$AU: gas in blue (identical values in both plots), total dust (for $a_{\max} = 1\,$mm) in orange and individual dust species used for modelling the wind in grey, according to Sect.~\ref{sec:Methods:rho:base}.
    \textit{Left:} Results for a globally `fixed' dust-to-gas ratio of 0.01, and \textit{right:} for the `variable' dust scale height dependent on $a_0$.
    The disk-wind interface (green dotted line) is characterised by a sharp drop-off of both gas and dust densities.
    This drop-off is particularly pronounced for the `fixed' total dust density at the disk surface, since we have no grains $> 11.5\,\mu$m in the wind.}
    \label{fig:base-densities}
\end{figure*}

The resulting base densities for the `fixed' and `variable' setups are shown in Fig.~\ref{fig:base-densities} for $R = 20\,$AU (with $R=\sqrt{x^2+y^2}$), the approximate value from which the largest grains are entrained (see Paper I).
For the same underlying gas density profile, we see a clear difference in dust scale heights between the `fixed' and `variable' models, especially for the larger $a_0$.
Furthermore, as the density profiles of the eight $a_0$ happen to be comparable in magnitude for the `fixed' setup, its mass contributions of the grain size bins are similar.

\subsubsection{Dust densities in the wind}
\label{sec:Methods:rho:combine}

\begin{figure}
    \centering
    \includegraphics[width=\columnwidth,trim={25mm 109mm 33mm 133mm},clip]{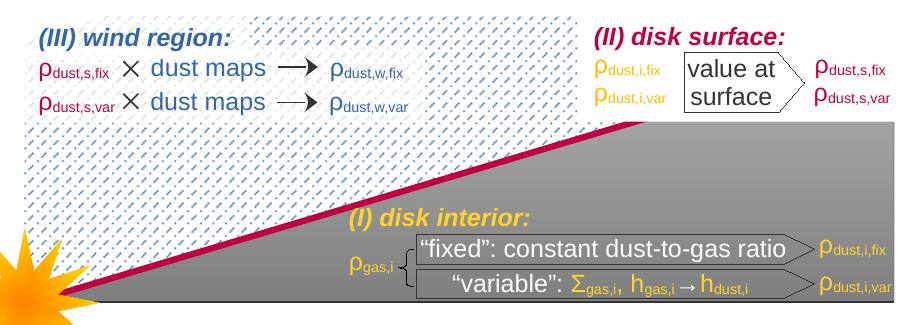}
    \caption{Schematics for determining $\varrho_\mathrm{dust}$ in the wind region:
    The dust densities in the disk are computed from $\varrho_\mathrm{gas}$ (see Sect.~\ref{sec:Methods:rho:base}).
    $\varrho_\mathrm{dust}$ at the disk surface is directly taken from these; it is then combined with the density maps from Sect.~\ref{sec:Methods:rho:traj-dist} to obtain $\varrho_\mathrm{dust}$ in the wind (see Sect.~\ref{sec:Methods:rho:combine}).}
    \label{fig:methods}
\end{figure}

For the dust densities within the disk, we directly used those computed in Sect.~\ref{sec:Methods:rho:base}; for the wind region, we combined the densities at the disk surface with the dust maps from Sect.~\ref{sec:Methods:rho:traj-dist}.
For clarity, the whole process is sketched in Fig.~\ref{fig:methods}.

Since the dust grids were computed for the wind region, the base densities need to be extracted from the position of the disk surface which is already affected by the wind; this corresponds to the minimum value along the density drop at the disk surface seen in Fig.~\ref{fig:base-densities}.
This approach also mostly reproduces the results of \citet{Booth-2021}, who find the rate of dust flux to gas flux across the ionisation front to be $\lesssim 1$ for a setup similar to ours.

The dust densities in the wind region were slightly smoothed with a Gaussian filter ($\sigma = 2\,$AU) in order to smear out numerical artefacts in the form of very narrow, overdense outflow channels next to sparsely populated areas, caused by employing Lagrangian particles on top of a Eulerian grid.
This affected neither the total dust masses in the wind, nor the results of the radiative-transfer computations.

The `fixed' setup represents the highest (maximum) base densities possible, and the `variable' one produces a more realistic estimate.
In order to provide a minimum case as well -- and to allow for a better estimation of the visibility of XEUV winds -- we further added models with a fully dust-free wind region (`no wind') for both setups.

\subsubsection{Dust mass-loss rates}
\label{sec:Methods:rho:Mdot}

From the individual dust trajectories, we also know the velocities at which the dust travels when leaving the computational domain at $r \simeq 300\,$AU.\footnote{The dust grains reach escape velocity before leaving the domain, see Paper I.}
In addition, we found a very clear correlation between the launching point of a grain along the disk surface and its final velocity; having a mean velocity per radially outermost $(r, \vartheta)$-bin and knowing its volume (assuming azimuthal symmetry), this allows for the computation of the wind mass-loss rates for each $a_0$.
Simple numerical integration was then employed to obtain a total dust mass-loss rate.

\subsection{Synthetic observations}
\label{sec:Methods:obs}

In order to get a handle on the observability of the dust outflows of our XEUV wind model, we produced synthetic images for inclinations of $i \in \lbrace 0; 30; 60; 75; 90 \rbrace^\circ$, using \texttt{RadMC-3D} \citep{Dullemond-2012}.\footnote{\texttt{RadMC-3D}: \href{http://www.ita.uni-heidelberg.de/~dullemond/software/radmc-3d/}{[link]}. Version 2.0 was used for this work.}
These were further converted into simulated instrument responses for the James Webb Space Telescope (JWST) Near Infrared Camera (NIRCam) scattered-light imaging \citep{Rieke-2003, Rieke-2005} using \texttt{Mirage} \citep{Hilbert-2019},\footnote{\texttt{Mirage}: \href{https://mirage-data-simulator.readthedocs.io/}{[link]}. Version 2.1.0 was used for this work.} and the JWST pipeline,\footnote{\texttt{jwst}: \href{https://jwst-pipeline.readthedocs.io/}{[link]}} and polarised-light imaging with the Spectro-Polarimetric High-contrast Exoplanet REsearch (SPHERE) + InfraRed Dual-band Imager and Spectrograph (IRDIS) instrument of the Very Large Telescope (VLT) \citep{Beuzit-2019}.\footnote{Overview of the available filters: \href{https://jwst-docs.stsci.edu/near-infrared-camera/nircam-instrumentation/nircam-filters}{[JWST NIRCam]}, \href{https://www.eso.org/sci/facilities/paranal/instruments/sphere/inst/filters.html}{[SPHERE]}}

\subsubsection{Dust opacities}
\label{sec:Methods:obs:opacities}

We employed the \texttt{dsharp\_opac} package \citep{Birnstiel-2018} to compute two sets of opacities:\footnote{\texttt{dsharp\_opac}: \href{https://github.com/birnstiel/dsharp_opac/}{[link]}}
Firstly, Disk Substructures at High Angular Resolution Project (DSHARP) opacities, employing optical constants from \citet[][for water ice]{Warren-2008}, \citet[][for silicates]{Draine-2003c}, and \citet[][for troilite and organics]{Henning-1996}.
The overall material density for this mix was combined with a porosity of $\simeq 0.4$ to maintain the internal grain density of $\varrho_\mathrm{grain} = 1.0\,\mathrm{g/cm}^3$ from Paper I.
Secondly, astrosilicate opacities, also using \texttt{dsharp\_opac} with $\varrho_\mathrm{grain} = 1.0\,\mathrm{g/cm}^3$.

The \texttt{RadMC-3D} runs revealed dust temperatures $T_\mathrm{dust} < 115\,$K for $r > 35\,$AU in the wind region; so the wind may be too hot to contain (water) ice.
Additionally, the stellar radiation field may further photo-desorb any ices.
Thus we used the DSHARP opacities for the disk region and pure astrosilicate opacities for the wind region of the dust density maps; this will slightly enhance the visibility of dust grains in the wind, which serves our intention of providing a best-case scenario for their observability.

\subsubsection{Radiative transfer}
\label{sec:Methods:obs:radmc}

The dust density maps were expanded to full 3D models by assuming midplane and azimuthal symmetry.
Our grids were illuminated by the star detailed in Sect.~\ref{sec:Methods:rho}, modelled as a simple black body with $R_* = 2.5\,R_\odot$ and $T_* = 5000\,$K,\footnote{We checked that our results are not affected by changing $T_*$ to 4000K.} and the radiative transfer was performed with 400 logarithmically spaced wavelength points in the interval $10^{-1} \leq \lambda\,[\mu\mathrm{m}] \leq 10^{4}$.
Using the full anisotropic scattering of \texttt{RadMC-3D}, we created SEDs, and also $I$, $Q$, and $U$ images at a resolution of $800 \times 800$ pixels (i.e. $>1$ pixel per AU).

As wavelengths $\lambda_\mathrm{obs}$ for the simulated images, we chose $0.7\,\mu$m (for JWST NIRCam's F070W filter), $1.2\,\mu$m (for JWST NIRCam's F115W filter, and SPHERE+IRDIS's $J$-band), $1.6\,\mu$m (for JWST NIRCam's F150W, F150W2+F162M and F150W2+F164N filters, and SPHERE+IRDIS's $H$-band), and $1.8\,\mu$m (for JWST NIRCam's F182M filter in combination with the MASRK210R coronagraph).
Additionally, we investigated $\lambda_\mathrm{obs} \in \lbrace 0.4, 3.2 \rbrace \,\mu$m.

For the results shown in this work, we used $N_\mathrm{phot}^\mathrm{therm} = 10^8$ photon packets for the Monte-Carlo computation of the dust temperature maps, and $N_\mathrm{phot}^\mathrm{scat} = 10^6$ photon packets for the imaging.
We furthermore checked that for the results with the most prominent wind signature, these do not significantly change (a) for $N_\mathrm{phot}^\mathrm{scat} = 10^7$, and (b) when re-mapping the dust densities to a grid logarithmic in $r$ (and thus providing better resolution close to the star).\footnote{For an analysis of the amount of photon packets needed for viable results in \texttt{RadMC-3D}, see \citet{Kataoka-2015}.}

\subsubsection{Scattered-light instrument response for JWST NIRCam}
\label{sec:Methods:obs:sca}

JWST NIRCam will provide both high sensitivity and high angular resolution for upcoming observations.
If this instrument is able to pick up an outflow signature from a dusty XEUV wind, then observational probes into the existence of XEUV winds (as modelled here) would become possible.

Using \texttt{Mirage}, we synthesised a scattered-light instrument response for the wavelengths and filters listed in Sect.~\ref{sec:Methods:obs:radmc}.
As we find in Sect.~\ref{sec:Results} that the dust in the wind should be more visible at shorter wavelengths, we forwent the integration of an additional long-channel filter.

The protoplanetary disk was assumed to be located at a distance of 100\,pc; this resulted in a rather bright (probably overexposed) region around the star for lower $i$.
We ran synthetic imaging for the \texttt{SUB320} subarray of module B1, using ten integrations with ten groups each for various readout patterns (RAPID, MEDIUM8, DEEP8, etc).\footnote{The Astronomer's Proposal Tool (\texttt{APT}, \href{https://www.stsci.edu/scientific-community/software/astronomers-proposal-tool-apt/}{[link]}) gives science durations between 107\,s (RAPID) and 2010\,s (DEEP8) for these parameters.}
The resulting uncalibrated images were post-processed using the \texttt{jwst} pipeline.

As shown by \citet[][their Fig.~6]{Beichman-2010}, the JWST NIRCam coronagraphs provide a contrast $\gtrsim 10^{-4}$ for separations $\lesssim 0{\farcs}3$ (i.e. 30\,AU at 100\,pc), and are available for filters corresponding to $\lambda_\mathrm{obs} \gtrsim 1.8 \,\mu$m.
\texttt{Mirage} does not yet support full coronagraphic imaging simulations, so we only mimicked the effect of a coronagraph by applying an intensity reduction to the RadMC results, according to the instrument transmission, before processing them with \texttt{Mirage}.\footnote{The transmission data have been retrieved from \href{https://jwst-docs.stsci.edu/near-infrared-camera/nircam-instrumentation/nircam-coronagraphic-occulting-masks-and-lyot-stops}{[here]}: \href{https://jwst-docs.stsci.edu/files/97978137/97978146/1/1596073154569/transmissions.tar}{[file]}}$^\mathrm{,}$\footnote{The science duration for coronagraphic imaging is higher, ranging from 2095\,s (RAPID) to 39350\,s (DEEP8).}

\subsubsection{Polarised-light instrument response for SPHERE+IRDIS}
\label{sec:Methods:obs:pol}

The Stokes images $Q(\vec{r})$ and $U(\vec{r})$ emergent from \texttt{RadMC-3D}, in native resolution, can be used to predict the instrumental response in a polarisation observation with SPHERE+IRDIS.
As in Sect.~\ref{sec:Methods:obs:sca}, a distance of 100\,pc to the hypothetical object was assumed for this.

A convolution with a Gaussian kernel, $b_\mathrm{diff}$, whose dispersion $\sigma_\mathrm{diff}$ is set at the diffraction limit, $\sigma_\mathrm{diff} = \frac{1.2}{2\sqrt{2*\ln(2)}} \frac{\lambda}{D}$, represents an ideal adaptive-optics correction for a telescope of diameter $D$.
The IRDIS coronagraph is taken as a pill-box with diameter 0{\farcs}25, $T(\vec{r})$, and the diffracted Stokes images are thus approximated as $Q_s = (b_\mathrm{diff} \cdot Q) \times (b_\mathrm{diff} \cdot T)$ and $U_s = (b_\mathrm{diff} \cdot U) \times (b_\mathrm{diff} \cdot T)$.

An estimate of the polarised intensity $P = \sqrt{Q^2 +U^2}$, alleviated from the positive-definite bias, can be obtained with a linear combination of $Q_s$ and $U_s$ in the
so-called radial-Stokes formalism \citep{Schmid-2006}.
The idea, adapted to axially symmetric sources, is to assume that the polarisation direction is azimuthal, as in the case of single scattering \citep{Schmid-2006, Avenhaus-2014b, Garufi-2014, Canovas-2015, Avenhaus-2017, Avenhaus-2018, Monnier-2019}.
Here we follow the same convention as in \citet{Avenhaus-2018}, with
\begin{eqnarray}
    Q_\phi & =  & Q_s  \cos(2\,\phi) + U_s \sin(2\,\phi), \\
    U_\phi & = & -Q_s \sin(2\,\phi) + U_s \cos(2\,\phi),
\end{eqnarray}
where $\phi = \arctan(-\Delta\alpha/\Delta \delta)$ is the position angle and $\Delta\alpha$, $\Delta \delta$ are offsets along right-ascension and declination.
Another application of these radial Stokes parameters can be found in \citet{Casassus-2018}.

The $Q_\phi(\vec{r})$ image approximates the polarised intensity field in the case of perfectly azimuthal polarisation, that is when $U_\phi(\vec{r}) \equiv 0$.
However, as discussed by \citet{Canovas-2015}, the emergent radiation from an intrinsically axially symmetric object such as a protoplanetary disk, when seen at even moderate inclinations, undergoes multiple scattering events that produce a radial polarisation component, with a non-vanishing $U_\phi$ and negatives in $Q_\phi$.
The radial Stokes parameters are nonetheless widely used in the field, as they convey the same information as $Q$ and $U$, and we therefore estimate instrumental responses using this formalism.

In adaptive-optics-assisted imaging, it is often the case that an unresolved and very strong central signal, due for example to an inner disk or to net stellar polarisation, is spread out by the PSF wings to large stellocentric separations.
Since we use a Gaussian PSF to estimate the instrumental response, this effect should be negligible in the bulk of the disk in the synthetic images, but might be important near the edges of the synthetic coronagraph.
For consistency with the procedure applied to actual data, we also implemented the correction for `stellar polarisation', which refers to the subtraction of the large-scale pattern due to the convolution of the unresolved central component with the PSF.
For a synthetic `stellar polarisation' subtraction, we simply define a radius, in polar coordinates $R=\sqrt{\Delta \alpha^2 + \Delta \delta^2}$, that encloses all of the `stellar polarisation' signal, and set $Q^\star=Q$ and $Q^\star(R > R^\star)=0$, and similarly for $U^\star$.
In practice, we chose the same radius for $R^\star$ as that of the synthetic coronagraph, or $R^\star=0\farcs125$.
We then applied the same radial Stokes formalism to $Q^\star$ and $U^\star$ to produce $Q^\star_\phi$ and $U^\star_\phi$, which were then subtracted from the predicted $Q_\phi$ and $U_\phi$ images.
The stellar polarisation subtraction had but a small impact on the resulting images.

The IRDIS observations of DoAr\,44 in $H$-band, presented in \citet{Avenhaus-2018}, can be compared with the RT predictions in \citet{Casassus-2018} to estimate the expected instrumental response for similar targets, that is other T-Tauri stars such as considered in this work.
We calibrated the observed $Q_\phi$ and $U_\phi$ images of DoAr\,44 by scaling with the RT predictions, thus extracting the noise level in $H$-band.

The synthetic images of the emergent polarisation in this work were thus cast into the radial-Stokes formalism, and reinterpolated to match the IRDIS CCD array, with the addition of Gaussian noise.
This ensures that the predicted instrumental response represents a concrete setup in realistic conditions.

\subsection{Limitations of the model}
\label{sec:Methods:limits}

In order to limit computational costs, we made a number of simplifying assumptions.
Firstly, we neglected dust sublimation.
The inner boundary of the computational domain is placed at $r = 0.33\,$AU; this allows us to capture the full extent of the photoevaporative wind \citep{Picogna-2019}.
This inner boundary also is well beyond the sublimation radius; assuming a dust sublimation temperature $T_\mathrm{sub} \approx 1500\,$K \citep{Pollack-1994, Muzerolle-2003, Robitaille-2006, Vinkovic-2009}, \texttt{RadMC-3D} yields $T_\mathrm{dust} \ll T_\mathrm{sub}$ for $r > 0.2\,$AU in the wind regions of our models.

Secondly, radiation pressure for grain acceleration was not included in our model (see Paper I).
Recently, \citet{Owen-2019} have argued that radiation pressure may drive the bulk of the dust mass loss if grains are fragmented to small enough sizes ($a_0 \lesssim 0.6\,\mu$m for our setup, see Paper I); and \citet{Vinkovic-2021} have shown that radiation pressure may severely affect dust trajectories in a magneto-hydrodynamic (MHD) wind region within $r \lesssim 30\,R_*$.
However, almost all of our smaller grains are entrained from the disk surface anyways (see Table~\ref{tab:particle-counts}), implying that additional radiation pressure should be negligible for our setup (the fraction of entrained grains could not be much higher anyways).
Furthermore, radiation pressure could lead to a small speed-up of the grains in the wind region; this would then lead to slightly reduced densities there as the grains would be blown out faster.
Yet \citet{Booth-2021} have argued that radiation pressure should not strongly affect grain entrainment by (X)EUV winds at least for an advection-dominated scenario.


\section{Results}
\label{sec:Results}

\subsection{Dust distribution in the wind}
\label{sec:Results:dist}

\subsubsection{Dust densities}
\label{sec:Results:dist:rho}

\begin{figure*}
    \centering
    \includegraphics[width=0.99\textwidth]{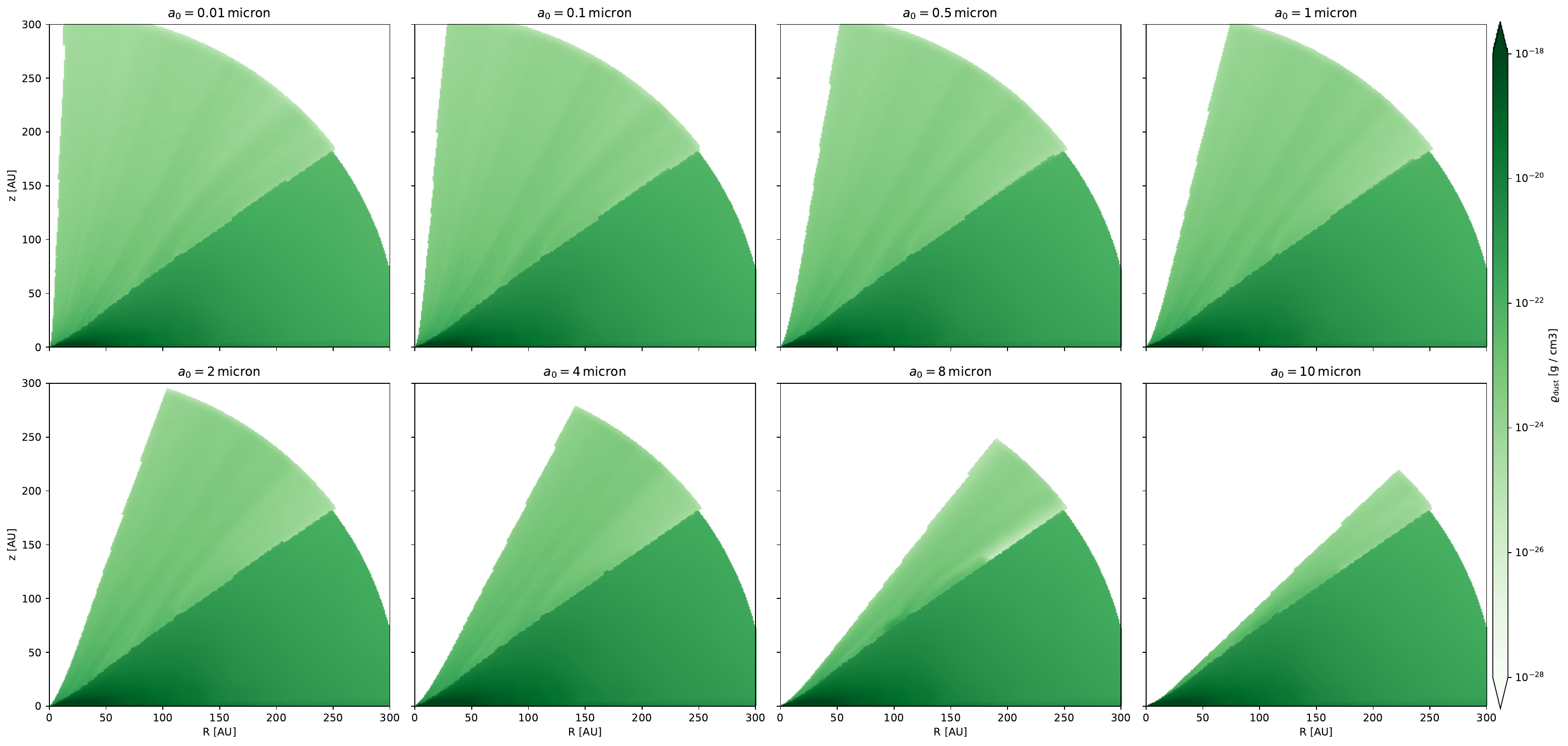}
    \caption{Dust densities for the XEUV-irradiated primordial disk in $(R, z)$ for a `fixed' dust-to-gas ratio throughout the disk.
    Enough material of all grain sizes is present at the base of the XEUV-driven outflow, resulting in high dust densities in the wind.
    Due to the model setup, the disk-wind interface is clearly discernible.}
    \label{fig:dust-densities-fixed}
    \includegraphics[width=0.99\textwidth]{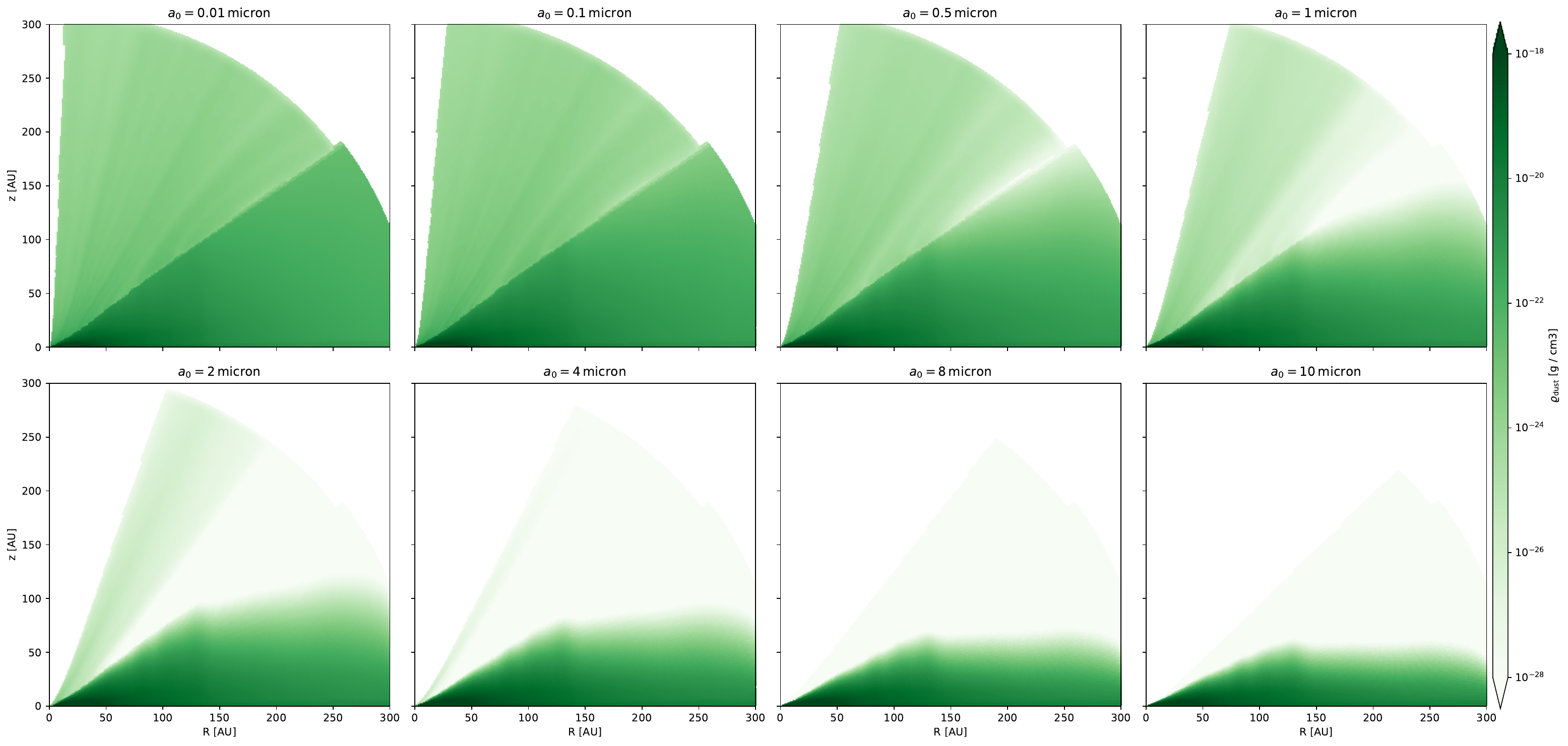}
    \caption{As Fig.~\ref{fig:dust-densities-fixed}, but for a `variable' dust-to-gas ratio in the disk; same grain sizes and colour bar.
    The (lower) `variable' dust scale height results in much lower dust densities in the wind than in the `fixed' model, even for very small grains.
    The slightly wiggly appearance of the in-disk densities at $R \simeq 150\,$AU is caused by the underlying gas disk; it can be ignored as most dust entrainment occurs for smaller $R$ (in respect to both total mass and radius).}
    \label{fig:dust-densities-variable}
\end{figure*}

The dust density maps in $(R, z)$ for the eight $a_0$ modelled are shown in Figs.~\ref{fig:dust-densities-fixed} (`fixed') and \ref{fig:dust-densities-variable} (`variable').
The `fixed' setup entails relatively high dust densities in the wind region for all $a_0$, albeit with bigger grains reaching lower maximum $z$ in the wind for a given $R$ (as expected from Paper I).
The disk surface, that is the transition region from the very smooth-looking disk regions to the outflow-dominated wind regions, is clearly visible for all $a_0$.
It marks a density decrease of about two orders of magnitude, as expected from Fig.~\ref{fig:base-densities}.

For the `variable' setup, the high-density regions for larger $a_0$ are compressed towards the disk midplane, and the dust content of the upper disk layers is strongly reduced.
This leads to a wind region much less populated by dust grains of $\mu$m size; the very light green regions in Fig.~\ref{fig:dust-densities-variable} indicate $\varrho_\mathrm{dust} < 10^{-28}\,\mathrm{g/cm}^{3}$.

\begin{figure}
    \centering
    \includegraphics[width=\columnwidth]{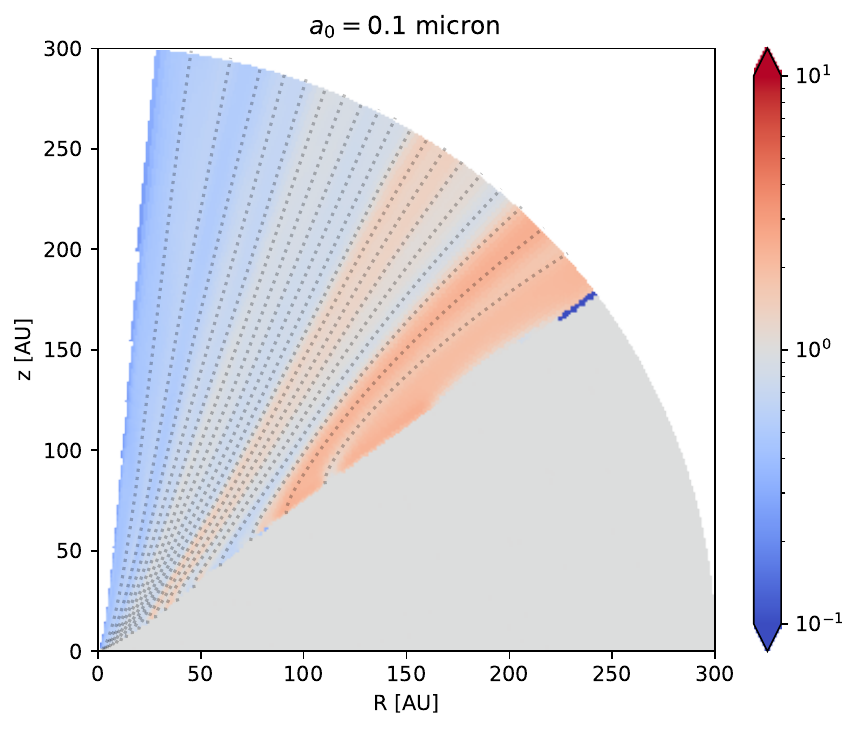}
    \caption{Ratio between gas density and dust density for $a_0 = 0.1\,\mu$m for the `fixed' model, normalised to the value in the disk for clarity.
    The density of this small, and hence well-coupled grain species closely follows the gas density.
    For comparison with the density variations in the dust, the gas streamlines (one per 5\% $\dot{M}_\mathrm{gas}$) are added as dotted grey lines.}
    \label{fig:dust-densities-dtg}
\end{figure}

As expected, especially small dust grains are well-coupled to the gas.
Fig.~\ref{fig:dust-densities-dtg} shows the deviation between the dust-to-gas ratio computed according to Sect.~\ref{sec:Methods:rho:base} and the ratio yielded by our `fixed' model for small grains ($a_0 = 0.1\,\mu$m).
Simply put, this illustrates the deviation of the dust densities we computed, from a model assuming a constant dust-to-gas ratio for the wind region as well.
The gas streamlines are included as dotted grey lines; the directions of the gas flow and the dust outflow channels match well, as would be expected (see Paper I).
This, in turn, validates our approach of using the dust densities at the wind-dominated side of the disk surface (see Sect.~\ref{sec:Methods:rho:combine}).

For wind-driven outflows originating from $R \lesssim 10\,$AU, we notice a slight decrease of the relative dust content.
This matches the results of \citet{Hutchison-2016b, Hutchison-2021}, who find a decrease of the dust-to-gas ratio in the wind region in their models.
Conversely, we see enhanced densities close to the disk surface far from the star ($R \gtrsim 100\,$AU).
These are caused by grains being picked up by the wind at large $R$, and then travelling only slightly above the disk surface at a comparably low speed.
This density enhancement occurs for all grain sizes investigated.
Conversely, in the `variable' model, the relative dust-to-gas ratios in the wind are smaller than for the `fixed' setup, and $< 1$ for $a_0 \geq 0.5\,\mu$m.

Interestingly, in Fig.~\ref{fig:dust-densities-dtg}, the uppermost outflow channel with slightly enhanced dust densities starts from $R \approx 20\,$AU (also somewhat visible in Figs.~\ref{fig:dust-densities-fixed} and \ref{fig:dust-densities-variable}), which coincides with the $R$-value from which we found the largest grains to be entrained in Paper I.
This shows a correlation between general wind strength, material outflow and maximum entrained grain size.

\subsubsection{Dust mass-loss rates}
\label{sec:Results:dist:Mdot}

\begin{figure*}
    \centering
    \includegraphics[width=0.495\textwidth]{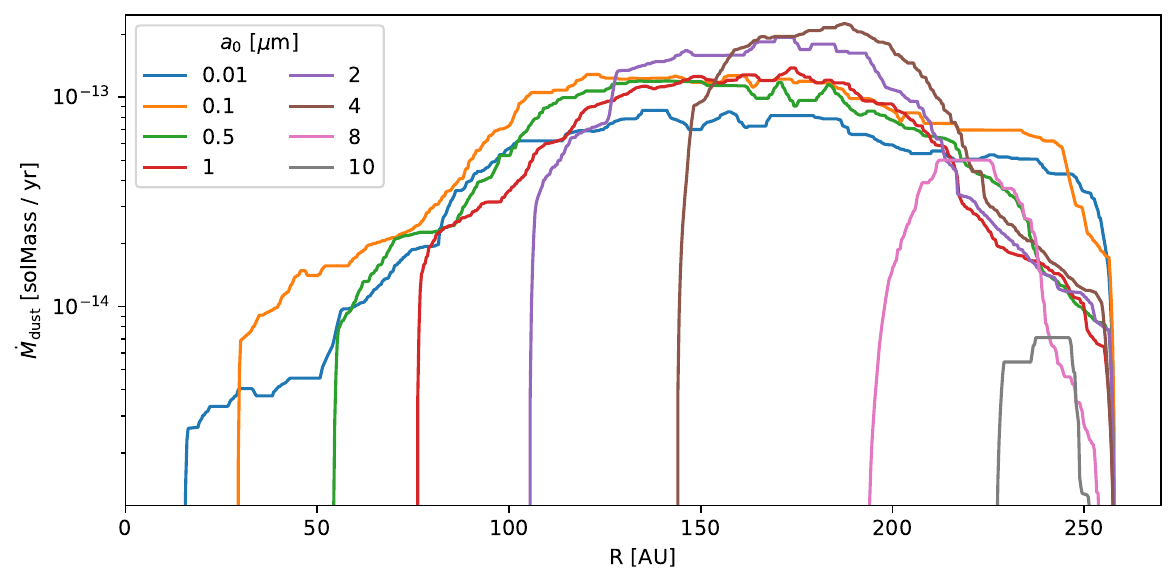}
    \includegraphics[width=0.495\textwidth]{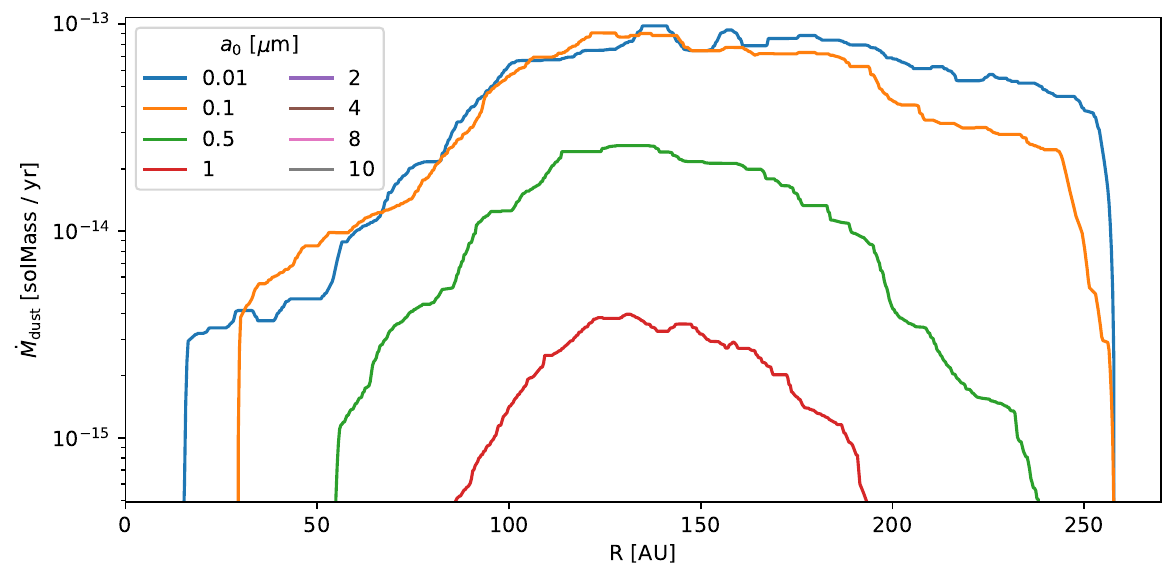}
    \caption{Median-filtered (for clarity) $\dot{M}_\mathrm{dust}$ over $R$ for the `fixed' (\textit{right}) and `variable' (\textit{left}) scenarios, accounting for an assumed midplane symmetry.
    For no grain size do we find high $\dot{M}_\mathrm{dust}$ in the jet region.}
    \label{fig:dust-massloss}
\end{figure*}

The XEUV-induced gas mass-loss rate is $\dot{M}_\mathrm{gas} \simeq 3.7 \cdot 10^{-8}\,\mathrm{M}_\odot$/yr (see Sect.~\ref{sec:Methods:rho}).
The dust mass-loss rates $\dot{M}_\mathrm{dust}$ per $a_0$ for the `fixed' and `variable' cases are shown in Fig.~\ref{fig:dust-massloss}; the corresponding (cumulative) values are listed in Table~\ref{tab:dust-massloss}.

\begin{table}
    \centering
    \caption{Total $\dot{M}_\mathrm{dust}$ per grain size per model and cumulative values.}
    \def\arraystretch{1.2}
    \begin{tabular}{ccc}
        \hline \hline
        & \multicolumn{2}{c}{$\dot{M}_\mathrm{dust}$ [M$_\odot$/yr]} \\
        $a_0$ & `fix' & `var' \\ \hline
        0.01 & 5.5e-12 & 5.8e-12 \\
        0.1 & 8.1e-12 & 4.6e-12 \\
        0.5 & 6.8e-12 & 9.9e-13 \\
        1 & 6.0e-12 & 1.1e-13 \\
        2 & 7.3e-12 & 7.1e-15 \\
        4 & 6.6e-12 & 3.6e-17 \\
        8 & 1.0e-12 & 3.6e-22 \\
        10 & 1.9e-13 & 9.5e-30 \\ \hline
        (sum) & 4.1e-11 & 1.2e-11\\ \hline
    \end{tabular}
    \label{tab:dust-massloss}
\end{table}

For the `fixed' setup, the dust mass-loss rate is fuelled by grains of all sizes, with only the contribution from $a_0 \gtrsim 8\,\mu$m falling off; this is expected since it represents the underlying mass distribution (see Sect.~\ref{sec:Methods:rho:base}) and entrainment rates (see Table~\ref{tab:particle-counts}).
For all $a_0$, $\dot{M}_\mathrm{dust}$ peaks closer to the disk surface than to the jet region.
The cumulative mass-loss rate is $\dot{M}_\mathrm{dust} \simeq 4.1 \cdot 10^{-11}\,\mathrm{M_\odot/yr} \simeq 1.1 \cdot 10^{-3} \, \dot{M}_\mathrm{gas}$.
Coincidentally, the sum of the mass contributions from the MRN distribution for $a_0 < 11.5\,\mu$m is $\approx 11\%$; so accounting for the underlying dust-to-gas ratio of 0.01, this matches with the setup.
Of course, not all large grains are entrained; considering that the eight size bins we employed have approximately similar mass contributions in the disk, this hints at a possible over-estimate for the dust masses in the wind (which suits our purposes), most probably due to the convolution of base densities and dust maps described in Sect.~\ref{sec:Methods:rho:combine}.

The `variable' setup entails a lower $\dot{M}_\mathrm{dust}$, yielding a cumulative value of $\dot{M}_\mathrm{dust} \simeq 1.2 \cdot 10^{-11}\,\mathrm{M_\odot/yr} \simeq 3.2 \cdot 10^{-4} \, \dot{M}_\mathrm{gas}$, about one third of the values for the `fixed' scenario.
Here, the bulk of the dust mass-loss is due to the grains $< 1\,\mu$m.
The contribution from the smallest $a_0$ is even slightly higher than in the `fixed' case; this is most likely due to the vertical settling-mixing equilibrium employed favouring higher vertical scale heights for small grains.

\subsection{Synthetic observations}
\label{sec:Results:synth}

\begin{figure*}
    \centering
    \includegraphics[width=0.99\textwidth]{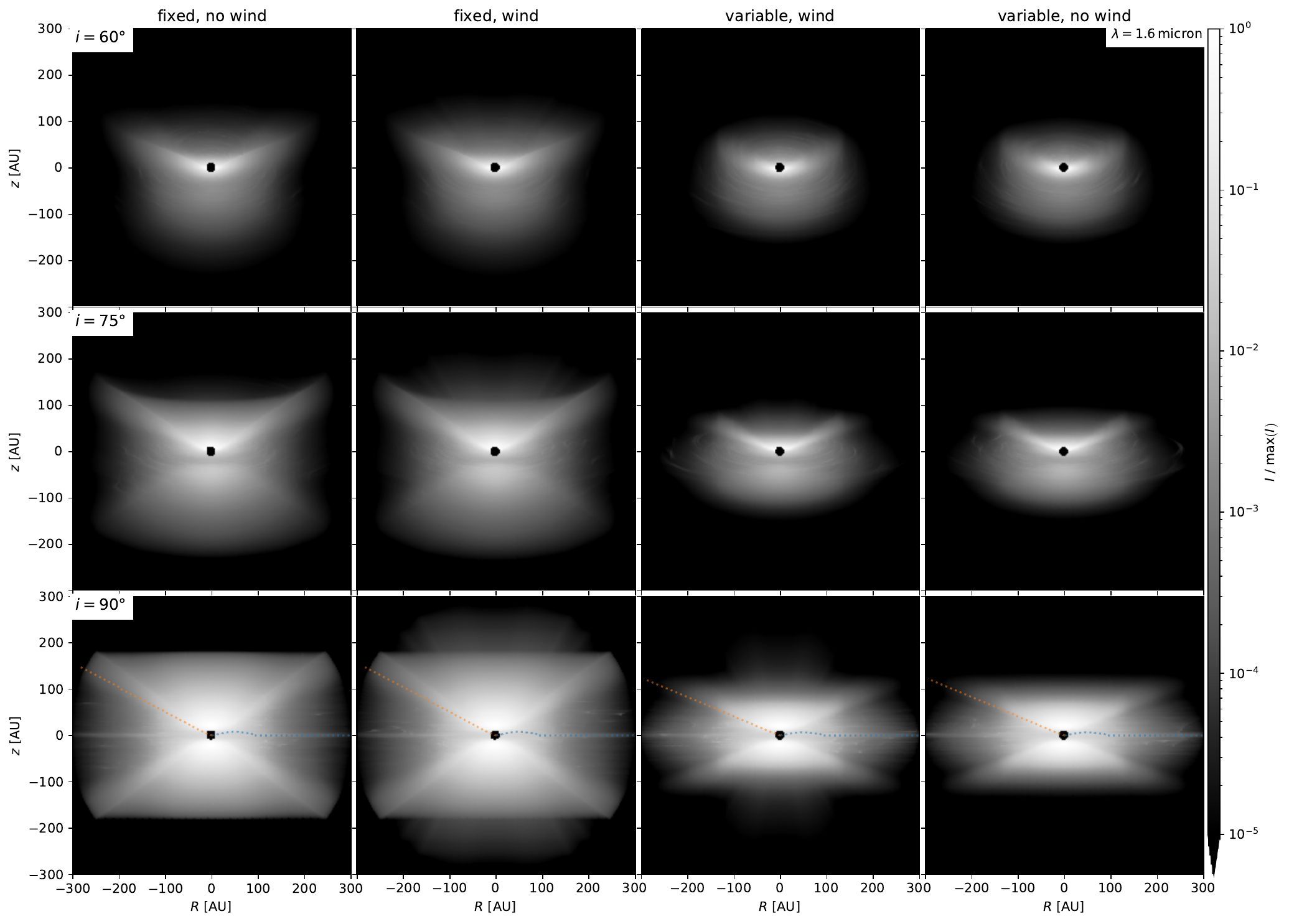}
    \caption{Radiative-transfer intensities for $\lambda_\mathrm{obs} = 1.6\,\mu$m.
    The $\max(I)$ for the logarithmic colour scale is taken for each plot individually, after application of an artificial coronagraph of radius 10\,AU.
    The $(\tau = 1)$-surfaces from $r=0$ and $z=\infty$ are indicated by dotted orange and blue lines, respectively.
    \textit{Rows:} inclinations $i \in \lbrace 60; 75; 90 \rbrace^\circ$, \textit{columns:} results without (first and fourth) and with (second and third) dusty wind outflow; `fixed' model on the \textit{left}, `variable' one on the \textit{right}.
    The wind signature is most noticeable at high inclinations, revealing a chimney-like structure with a distinctly narrower opening angle than the disk surface; it remains clearly less bright than said disk surface.
    The `variable' dust scale height yields a much fainter outflow signal.}
    \label{fig:radmc-1.6}
\end{figure*}

The scattered-light intensities simulated from the dust densities of Figs.~\ref{fig:dust-densities-fixed} and \ref{fig:dust-densities-variable} for $\lambda_\mathrm{obs} = 1.6\,\mu$m are shown in Fig.~\ref{fig:radmc-1.6}.
The `wind' images (inner columns) are accompanied by corresponding `no wind' images (outer columns) for direct comparability.
To provide better visibility of fainter features, an artificial coronagraph of $r=10\,$AU was introduced at the location of the star.

The differences between the `fixed' and `variable' models are much more pronounced than between their respective `wind' and `no wind' cases.
In all cases, the wind appears featureless at low inclinations $i$ ($i \lesssim 45^\circ$) and for hence more face-on disks, even despite using logarithmic stretch.

For the `fixed' setup (left columns of Fig.~\ref{fig:radmc-1.6}), cone- or chimney-like features start to envelop the $z$-axis for more edge-on objects ($i \gtrsim 60^\circ$); however, they are distinctly less bright than the disk component.
The relative intensity of the wind features is $I/I_{\max} \lesssim 10^{-4.5}$, $I_{\max} \equiv \max(I)$, at $i = 90^\circ$ (best case), rendering these wind features at least challenging to observationally detect.

For the `variable' model, the dusty outflow is less bright, and is noticeable at $i \gtrsim 75^\circ$; it is qualitatively similar to the `fixed' results.
Furthermore, due to the reduced dust scale heights, the disk appears flatter, which is also illustrated by the $(\tau=1)$-lines indicating where the optical depth reaches one starting from $r=0$ (orange lines in Fig.~\ref{fig:radmc-1.6}).
In addition, the disk appears smaller at $i<90^\circ$, caused by the decreased delivery of grains $\gtrsim 1\,\mu$m to the disk surface (see Fig.~\ref{fig:dust-densities-variable}).

The low dust densities in the outflow do not lead to a notable enhancement in apparent dust scale heights; again, this can be seen from the ($\tau = 1$)-surfaces for an observer at $z=\infty$ (blue lines in Fig.~\ref{fig:radmc-1.6}).
These surfaces do not visibly differ between the respective `wind' and `no wind' models.

\begin{figure*}
    \centering
    \includegraphics[width=0.99\textwidth]{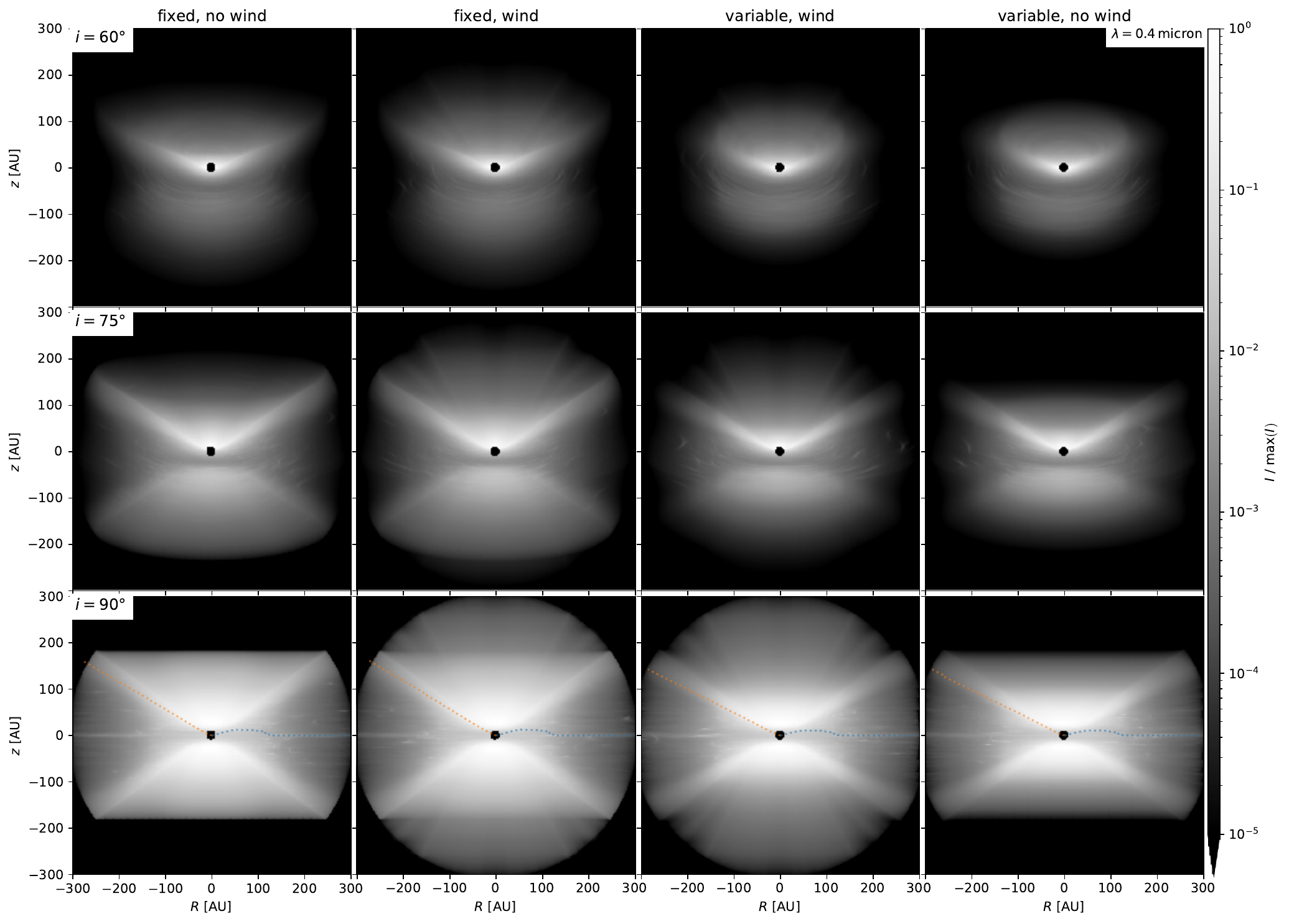}
    \caption{Radiative-transfer intensities for $\lambda_\mathrm{obs} = 0.4\,\mu$m, all else equal to Fig.~\ref{fig:radmc-1.6}.
    Both the `fixed' and `variable' models produce a noticeable wind signature (in logarithmic stretch), and the disks look quite similar.}
    \label{fig:radmc-0.4}
\end{figure*}

The scattered-light images for $\lambda_\mathrm{obs} = 0.4\,\mu\mathrm{m} = 400\,$nm are shown in Fig.~\ref{fig:radmc-0.4}.
For smaller $a_0$, the density maps for the `fixed' and `variable' models are more similar (see Figs.~\ref{fig:dust-densities-fixed} and \ref{fig:dust-densities-variable}); this results in more similar images at $\lambda_\mathrm{obs} = 0.4\,\mu$m compared to $1.6\,\mu$m.
Thus, the scattered-light intensities for the `no wind' cases are almost model-independent at $i=90^\circ$; at $i<90^\circ$, the `variable' disk still appears smaller.
Yet it looks considerably more puffed up than at $\lambda_\mathrm{obs} = 1.6\,\mu$m, as would be expected from a size-dependent vertical-settling prescription.

The XEUV-driven dust outflow signature is quite comparable for the `wind' models, and reaches relative intensities of $I/I_{\max} \lesssim 10^{-3.5}$ at $i = 90^\circ$.
The $(\tau = 1)$-surfaces are still almost identical within the `fixed' and `variable' models, meaning that the apparent dust scale height is still not significantly impacted.
Furthermore, the opening angle of the emerging cone feature(s) does not really change in comparison to $\lambda_\mathrm{obs} = 1.6\,\mu$m, indicating that it is unrelated to the maximum outflow heights parametrised in Paper I.

When comparing the SEDs from disks without and with a dusty wind, we found no features that clearly stood out above the Monte-Carlo-induced noise.
This indicates that the emission at the wavelength range investigated is dominated by the bound disk.

\subsubsection{Scattered-light images}
\label{sec:Results:synth:sca}

Due to the comparatively very bright inner area around the star, sporting radiative-transfer intensities of up to $\approx 2 \cdot 10^{8} \, \mathrm{nJy/arcsec}^2 \simeq 2 \cdot 10^{5} \, \mathrm{nJy/pix}$, JWST NIRCam images synthesised without accounting for a coronagraph are clearly over-exposed even using the RAPID readout pattern; this leads to an overexposure pattern which bleeds far into the disk.
The relative brightness of the dusty XEUV outflow is thus too low to be picked up, or contaminated by overflow from the inner region.
This means that we cannot use the F070W filter for our purposes.

\begin{figure*}
    \centering
    \includegraphics[width=0.99\textwidth]{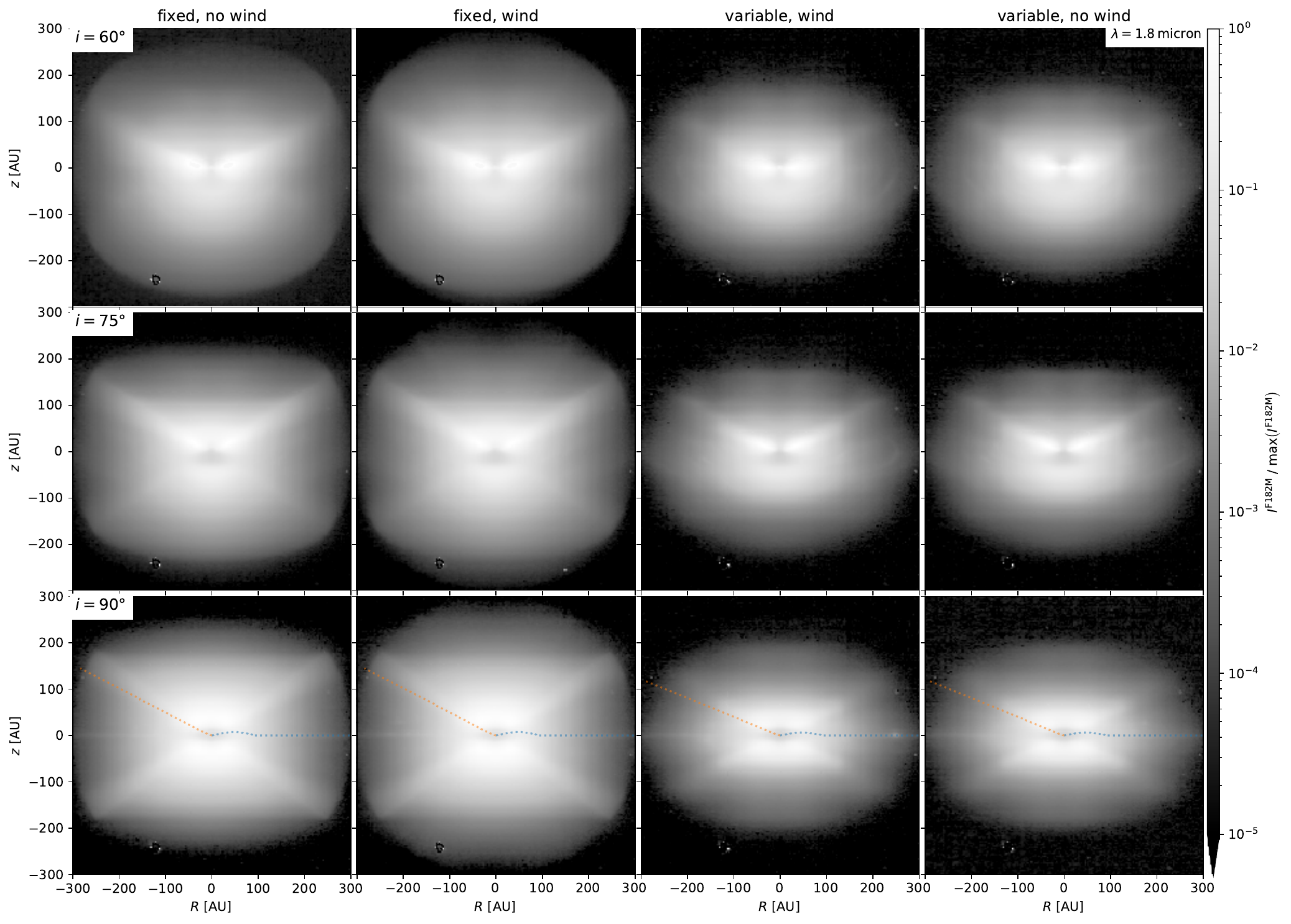}
    \caption{Synthesised intensities for JWST NIRCam's F182M filter with a simulated MASK210R coronagraph (MEDIUM8 readout pattern, 10 groups for 10 integrations).
    The orange and blue lines indicate the $(\tau = 1)$-surfaces from $r = 0$ and $z = \infty$.
    We find a slight difference in the vertical extent of the `no wind' and `wind' `fixed' models.
    At $i \gtrsim 75^\circ$, a faint cone-like pattern emerges, in agreement with the direct radiative-transfer results of Fig.~\ref{fig:radmc-1.6}.}
    \label{fig:nircam-1.8-mask}
\end{figure*}

Instead, we switched to $\lambda_\mathrm{obs} = 1.8\,\mu$m with the F182M filter in combination with the simplified coronagraph implementation as described in Sect.~\ref{sec:Methods:obs:sca}; the resulting synthesised images are shown in Fig.~\ref{fig:nircam-1.8-mask}.
For these, the MEDIUM8 readout pattern was employed, providing the best middle ground between an overexposed central region (DEEP8) and noisy outer regions (RAPID).
Varying the number of groups and integrations yielded very similar results.

For the `fixed' setup, we find a slightly enhanced vertical extent for $i \gtrsim 75^\circ$, reminiscent of the cone-shaped feature of the radiative-transfer results (see Fig.~\ref{fig:radmc-1.6}).
However, the cone itself does not stand out as clearly, and the corresponding difference between the `wind' and `no wind' images may be too small to be used as a definite outflow indicator without detailed modelling of the source.
The `wind' and `no wind' images of the `variable' model do not differ in any noticeable way.

\subsubsection{Polarised-light images}
\label{sec:Results:synth:pol}

\begin{figure*}
    \centering
    \includegraphics[width=0.99\textwidth]{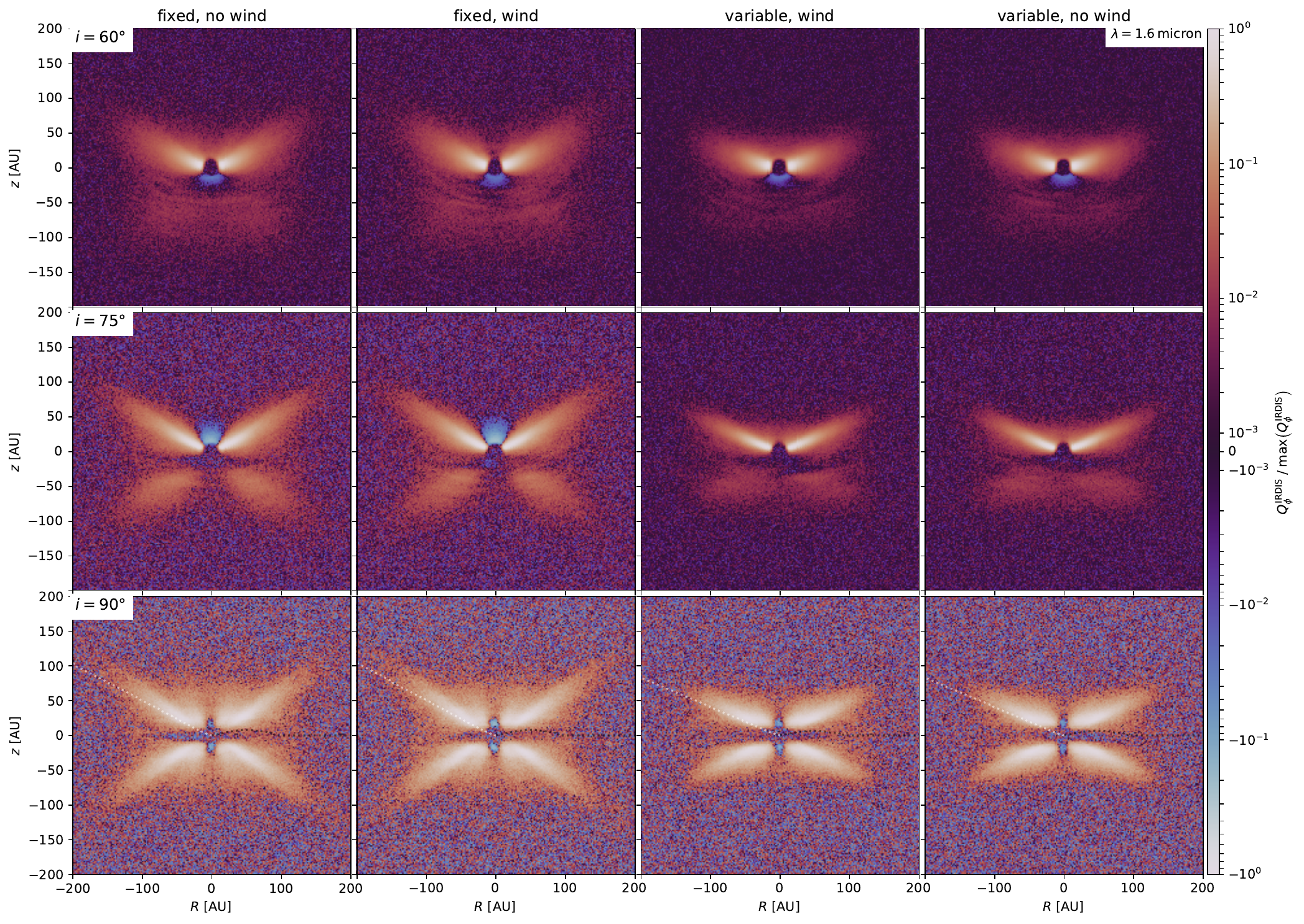}
    \caption{$Q_\phi$ maps (logarithmic stretch for the inner inner $200\,\mathrm{AU} \times 200\,\mathrm{AU}$) synthesised for SPHERE+IRDIS's $H$-band ($\lambda_\mathrm{obs} = 1.6\,\mu$m); $(\tau = 1)$-surfaces from $r=0$ and from $z=\infty$ added as dotted white and black lines, respectively.
    No distinct difference emerges between the disks without (first and fourth columns) and with (second and third columns) a dusty wind; the difference between the `fixed' and `variable' dust scale heights has a much larger impact on the synthesised instrument response.}
    \label{fig:irdis-1.6}
\end{figure*}

Fig.~\ref{fig:irdis-1.6} displays the $Q_\phi$ signal synthesised for SPHERE+IRDIS's $H$-band for the inclinations of interest.
There is no feature differentiating the dusty XEUV `wind' models from their `no wind' counterparts; this holds if we look at the data in logarithmic instead of linear stretch.
While differential imaging reveals a faint enhancement of the `variable' `wind' model compared to its `no wind' counterpart at $i=90^\circ$, the difference is too small to be useful for distinguishing between them.
This implies that, at least for the setup investigated here, XEUV-driven outflows may be too faint to be picked up by current instruments.
As can be seen from \citet[][their Fig.~3]{Avenhaus-2018}, SPHERE+IRDIS can detect a signal down to a relative intensity of $\gtrsim 10^{-5}$ under optimal conditions; however, we do not find significant wind features when looking at Fig.~\ref{fig:irdis-1.6} in logarithmic stretch, mainly because any possible signal disappears in the instrument noise.
Higher mass-loss rates, as for instance provided by a centrally concentrated MHD wind model, might provide a more distinct signal; this should be explored in future calculations.

Synthesised $J$-band images ($\lambda_\mathrm{obs} = 1.2\,\mu$m) do not exhibit any notable `wind' features either, and neither do images for $P$.
When looking at $Q_\phi$ maps for $\lambda_\mathrm{obs} = 0.4\,\mu$m without a synthesised instrument response, a wind signature starts to emerge for $Q_\phi / \max(Q_\phi) \lesssim 10^{-3}$ ($i=90^\circ$) in the same chimney-shaped region we have seen in Fig.~\ref{fig:radmc-0.4}.
This contrast is comparable to the one needed to see a wind signature in said image.


\section{Discussion}
\label{sec:Discussion}

Theoretical models of photoevaporative winds agree that qualitatively, a certain amount of dust is entrained in the outflow above the protoplanetary disk \citep[e.g.][]{Clarke-2001, Owen-2011a, Owen-2012a, Ercolano-2017b}.
However, to our knowledge, there has not yet been an estimate of the actual dust content of this outflow, and hence its observability.

\subsection{Dust distribution}
\label{sec:Discussion:dust}

\subsubsection{Dust densities}
\label{sec:Discussion:dust:rho}

The `hovering' (or at least slow-moving) dust grains seen for the `fixed' case in Fig.~\ref{fig:dust-densities-dtg} at larger $R$ might potentially cause a diffuse signal in scattered- and polarised-light images at high inclinations; the polarised-light images provided by SPHERE+IRDIS for RY~Lup \citep{Langlois-2018}, 2MASS~J16083070-3828268 \citep{Villenave-2019}, and MY~Lup \citep{Avenhaus-2018} could be examples of this.
As pointed out by \citet{Booth-2021}, the vertical transport mechanisms at play will heavily influence the vertical disk height, and thus the size of these grains.
The disk scale height itself would need to be better constrained via observational data sets for our models to be more refined.\footnote{Constraining the dust delivery at the base of the wind would require modelling the evolution of disk material, and self-consistently accounting for dust growth together with gas and dust hydrodynamics; this complex task is well beyond the scope of this work, where we take the approach of showing limiting cases.}
Nonetheless, our results seem to match the findings of \citet{Pinte-2008, Villenave-2020} and others, who find $\mu$m-sized dust grains well above the disk midplane; in addition, the presence of $\mu$m-sized dust grains at the disk surface (and probably above) is well-documented by early SED data \citep[][``small-grains problem'']{Dullemond-2005} and scattered-light images of edge-on disks.
Furthermore, the absence of big grains in the wind matches the results of \citet{Hutchison-2021, Booth-2021}, who find that the majority of grains in the wind has sizes well below the theoretically possible value; we also match their result that almost all grains delivered to the base of the outflow will be entrained (within a given Stokes number range).

\subsubsection{Mass-loss rates}
\label{sec:Discussion:dust:Mdot}

Due to $\dot{M}_\mathrm{dust} / \dot{M}_\mathrm{gas}$ being smaller than the assumed dust-to-gas ratio, the dust-to-gas ratio in the disk will increase as a result of photoevaporation, particularly in the regions where the latter is most efficient (i.e. around the gravitational radius).
Not accounting for probably varying rates and various other effects, the gas and dust masses in our model would be equal about 0.15\,Myr after XEUV photoevaporation starts.

This could potentially explain high dust-to-gas ratios as found for instance by \citet{Miotello-2017}.
It is also expected to favour planetesimal formation by the streaming instability; however, recent studies which assume a dust-free photoevaporative wind, disagree on the efficiency of this process \citep{Carrera-2017, Ercolano-2017c}.
Moreover, as found for instance by \citet{Kunitomo-2020}, photoevaporative winds are supposed to dominate over MHD outflows in the later stages of the disk lifetime, at least in terms of $\dot{M}_\mathrm{gas}$.
And indeed, dust entrainment may be significantly enhanced once the transition disk phase begins, that is when the midplane dust at the outer gap edge would automatically be located at the disk-wind interface without the need for efficient vertical transport mechanisms.

\subsection{Observability of XEUV winds}
\label{sec:Discussion:obs}

We find a chimney-like outflow signature in Figs.~\ref{fig:radmc-1.6} and \ref{fig:radmc-0.4}; however, MHD winds may well dominate in terms of dust entrainment in the immediate vicinity and the jet region of a star \citep[see e.g.][]{Miyake-2016}.
These MHD winds could even be prominent all the way down to the disk surface \citep{Rodenkirch-2020}.

\citet[][and sources therein]{Villenave-2020} have presented a collection of images of highly inclined disks in scattered light at $\mu$m-wavelengths ($0.4 \lesssim \lambda_\mathrm{obs} \,[\mu\mathrm{m}] \lesssim 2.2$).
Most of these objects (e.g. HH~30, 2MASS~J04202144+2813491, HV~Tau~C) do not exhibit well-defined cone-like features as those shown in Figs.~\ref{fig:radmc-1.6} and \ref{fig:radmc-0.4}.
Nonetheless, a diffuse emission above the disk surface is visible in some objects, and may be due to `hovering' dust grains. 
Some other objects \citep[e.g. IRAS~04158+2805, see][]{Villenave-2020} exhibit a distinct region of enhanced outflow above the disk midplane; yet this feature appears to extend to large scale heights above the disk midplane and involves the entire jet region.
Thus, it is quite difficult to trace the origin of this emission without a tailored modelling of the source.

The `variable' dust scale height entails a darkening of the outer disk regions ($R \gtrsim 140\,$AU).
This is probably caused by shadowing from the inner 140\,AU of the disk, after which the density especially of large grains around the disk surface drops due to vertical settling (see Fig.~\ref{fig:dust-densities-variable}).
In principle, this is comparable to the model `B' of \citet{Dullemond-2002}, which shows a puffed-up inner rim to produce a shadow.
It is not an effect caused by photoevaporative winds because the `wind' and `no wind' models are both affected.

\subsubsection{Predictions in scattered light}
\label{sec:Discussion:obs:sca}

Hubble Space Telescope (HST) images of disks have not yet shown any evidence for disk winds driven by internal photoevaporation; this may be because the signal is too faint,\footnote{External photoevaporation may have already been imaged, see \citet{ODell-1994, Miotello-2012}.} or the jet contaminates fainter features \citep{Wolff-2017}.
JWST NIRCam will provide notably higher contrast.
In theory, this means that as we have seen from Fig.~\ref{fig:nircam-1.8-mask}, the XEUV-driven dusty outflow modelled here may actually produce an observable signature, at least for the highly inclined disks of the `fixed' model.
Since the cone shape of the wind-driven dusty outflow does not stand out strongly, and provides only a slight increment in vertical height for the `wind' images over their `no wind' counterparts, it is debatable whether an isolated image of a primordial disk would suffice to determine the presence or absence of a wind.
Reducing (extending) the cutoff radius (here: $r \simeq 300\,$AU) of the model would likely enhance (degrade) the visibility of the outflow.

The `variable' setup does not exhibit any outflow pattern, despite the synthesised images showing faint traces of overexposure in their centre (which should enhance the visibility of far-out features).
On a different note, a higher underlying dust-to-gas ratio (see Sect.~\ref{sec:Methods:rho:base}) may lead to a more distinct wind pattern for the setup presented here (see also Dahlb{\"u}dding et al., in prep.).

The synthetic observations shown in this work have been computed from and for a primordial disk.
We conclude that intriguingly, with JWST NIRCam the observation of wind signatures is entering the realm of possibilities.

\subsubsection{Comparison to observations in polarised light}
\label{sec:Discussion:obs:pol}

The differential-polarised imaging data presented in \citet{Avenhaus-2018} and \citet{Garufi-2020} target 29 nearby T-Tauri stars, with stellar masses in the range $0.5 \leq M_*\,[\mathrm{M}_\odot] \leq 1.0$ \citep[][hereafter DARTTS-S sample]{Garufi-2020}, and correspond to the domain considered by our model predictions.
The DARTTS-S sample includes high-inclination sources, with a favourable view of material away from the disk midplanes.
In general, none of these T-Tauri stars display any evidence for nebulosity along the disk minor axis that might correspond to an XEUV wind; for instance, one can look at the $Q_\phi$ image of DoAr\,25 in $H$-band presented by \citet[][their Fig.~1]{Garufi-2020}.
The absence of conspicuous wind features is consistent with our predictions, and does not indicate the absence of XEUV-driven photoevaporative winds.

By contrast, MY\,Lup is the one object in the DARTTS-S sample with diffuse emission along the disk minor axis.
Some extended negatives along the disk minor axis in the data originally presented by \citet{Avenhaus-2018} prompted us to reduce the images using the IRDAP pipeline \citep{vanHolstein-2020}, with special attention to stellar polarisation subtraction; for this, the IRDIS images have been processed using an adaptive kernel smoothing technique called \texttt{denoise} \citep[part of the \texttt{splash} suite,][]{Price-2007}.\footnote{\texttt{denoise}: \href{https://github.com/danieljprice/denoise/}{[link]}}
The resulting data show some diffuse signal indicating a dusty outflow.
Furthermore, the signal is more conspicuous in $J$-band ($\lambda_\mathrm{obs} \approx 1.2\,\mu$m) than in $H$-band, which is consistent with Rayleigh-scattering if the entrained dust is smaller than $\sim 1\,\mu$m.
New polarisation observations at shorter wavelengths, for instance with SPHERE+ZIMPOL in the $V$-band broad-band filter, could confirm this tentative detection.

\begin{figure*}
    \centering
    \includegraphics[trim={5mm 0 30mm 15mm},clip,width=\textwidth]{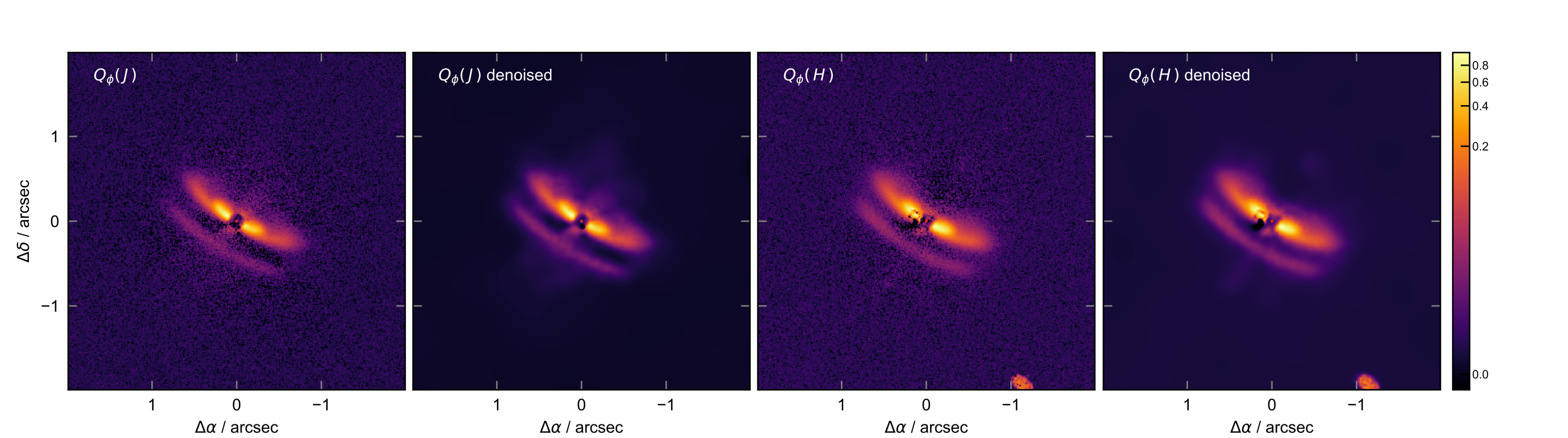}
    \caption{$Q_\phi$ images of MY\,Lup in $J$- and $H$-band, based on a new reduction of the data presented in \citet{Avenhaus-2018}.
    For both bands, the images are shown without and with smoothing by a denoising algorithm \citep{Price-2007}.
    There is a faint signal from the minor axis of the disk, which hints at a dusty outflow.}
    \label{fig:MYLup}    
\end{figure*}

As found by \citet{Alcala-2019}, MY\,Lup likely does not have a dust cavity, which would allow for an almost direct comparison to our primordial disk model.
Alternatively, they allow for the possibility that the high CO depletion found by \citet{Miotello-2017} could explain the high mass-accretion rates and enhanced winds \citep{Woelfer-2019}.
Furthermore, the presence of a dusty wind could explicate the anomalous extinction law towards the central star, which may have led to an overestimate of the age of the system \citep[$\approx 17\,$Myr, see][]{Alcala-2017, Alcala-2019}.

As seen in Fig.~\ref{fig:irdis-1.6}, we do not find a distinct signal enhancement along the disk minor axis for the $Q_\phi$ images synthesised for SPHERE+IRDIS from our `wind' models; we checked that this also holds for $\lambda_\mathrm{obs} = 1.2\,\mu$m.
Thus, while we would argue that the outflow seen in Fig.~\ref{fig:MYLup} is indeed driven by a wind, we cannot say whether said wind is an MHD wind, or possibly a CO-depleted, thus enhanced, XEUV one.
In any case, however, its corresponding $\dot{M}_\mathrm{dust}$ supposedly must be at least an order of magnitude higher than the values we have reported for our models ($\dot{M}_\mathrm{dust} \lesssim 4.1 \cdot 10^{-11}\,\mathrm{M_\odot/yr}$, see Sect.~\ref{sec:Results:dist:Mdot}), which would indicate a rapid depletion of dust at and around the disk surface.
This, in turn, would raise the question whether such a strong outflow could be stable for prolonged periods of time, or whether it may be a periodical or short-lived phenomenon.

So the possible detection of a dusty wind in MY\,Lup, while it is absent in the other DARTTS-S sources, suggests that our model is missing some of the diversity of real physical systems; more custom-tailored models for this source are needed in order to answer this question.
In addition, our findings lead us to expect dusty XEUV winds from primordial disks to be detectable at $0.4\,\mu$m (in logarithmic stretch), which is only a factor of $\sim 3$ in wavelength from the possible detection in MY\,Lup.


\section{Summary}
\label{sec:Summary}

Based on the trajectories of dust grains in the XEUV-driven wind region of a protoplanetary disk around a T-Tauri star, we have computed the dust density in said wind region.
Our main findings are as follows:
\begin{itemize}
    \item Photoevaporative winds can entrain $\mu$m-sized dust grains.
    
    \item The dust densities at the base of the disk wind heavily impact the dust content of the wind.
    
    \item The dust mass-loss rate due to XEUV winds is significantly lower than expected from the corresponding gas mass-loss rates.
    This may lead to an enhancement of the dust-to-gas ratio in the disk.
\end{itemize}

In addition, we have used these dust densities to synthesise observations in scattered light (for JWST NIRCam) and polarised light (for SPHERE+IRDIS).
The results have led us to the following conclusions:
\begin{itemize}
    \item Observations of dusty disk winds are challenging with current instrumentation -- which is limited to comparably large wavelengths --, even if the base of the wind is rich in $\mu$m-size grains.
    It is thus not surprising that current observational campaigns in scattered light have yet to find a definite wind signature.
    This does not necessarily imply the absence of a photoevaporative wind.
    
    \item Dusty winds launched from primordial disks should, in the majority of cases, not lead to a strong vertical puff-up of the disks.
    
    \item There is a tentative detection of a disk wind in MY Lup.
    This could be confirmed by tailored modelling of deeper observations, and observations at shorter wavelengths.
\end{itemize}

However, the models presented here are only applicable to primordial disks.
In a next step, we intend to investigate disk models with an inner cavity, which may produce more distinct signatures of XEUV-driven disk winds.


\begin{acknowledgements}

We would like to thank A.~Garufi, C.~Dullemond, P.~Rodenkirch, M.~Hutchison, and T.~Grassi for helpful discussions, and the (anonymous) referee for a constructive report which improved the manuscript.
\\
Furthermore, we thank C.~Dullemond for providing the soon-to-be-open-sourced \texttt{disklab} package together with T.B.
\\
This research was funded by the Deutsche Forschungsgemeinschaft (DFG, German Research Foundation), grant 325594231 (FOR 2634/1 and FOR 2634/2), and the Munich Institute for Astro- and Particle Physics (MIAPP) of the DFG cluster of excellence \textit{Origin and Structure of the Universe}.
B.E. and T.B. acknowledge funding by the Deutsche Forschungsgemeinschaft under Germany's Excellence Strategy -- EXC-2094-390783311.
S.C. and S.P. acknowledge support from Agencia Nacional de Investigaci{\'o}n y Desarrollo de Chile (ANID) through FONDECYT Regular grants 1211496 and 1191934.
T.B. acknowledges funding from the European Research Council (ERC) under the European Union's Horizon 2020 research and innovation programme under grant agreement No 714769.
The simulations have mostly been carried out on the computing facilities of the Computational Center for Particle- and Astrophysics (C2PAP).

\end{acknowledgements}


\bibliographystyle{aa}
\bibliography{Literature.bib}

\end{document}